\documentclass[aps,prd,onecolumn,eqsecnum,amsmath,nofootinbib,preprintnumbers,superscriptaddress]{revtex4-2}

\usepackage[dvipsnames]{xcolor}
\usepackage{color,graphicx,float,subfigure,xcolor}
\usepackage{amsfonts,amssymb,mathrsfs,times}
\usepackage{bm}
\usepackage{multirow}
\usepackage{mathtools}
\usepackage{dsfont}
\usepackage{setspace}
\usepackage{amsmath} 
\usepackage{graphicx}
\usepackage[colorlinks,linkcolor=blue,anchorcolor=blue,citecolor=blue]{hyperref}
\usepackage{ulem}
\usepackage{cancel}
\usepackage{subcaption}
\usepackage{amsthm}

\usepackage[justification=raggedright,singlelinecheck=false]{caption}

\begin{document}
\thispagestyle{empty}
\preprint{\hfill {\small {ICTS-USTC/PCFT-26-80}}}
%<<<<<<<<<<<<< TITLE >>>>>>>>>>>>>>>%
\title{Quasinormal modes of Kerr--de Sitter black holes in semi-open systems}
	
%<<<<<<<<<<<<< AUTHOR >>>>>>>>>>>>>>>%
%\author{$^b$}
%
% \email{}

\author{Ming-Fei Ji}
\email{jimingfei@mail.ustc.edu.cn}
\affiliation{Interdisciplinary Center for Theoretical Study and Department of Modern Physics,\\
University of Science and Technology of China, Hefei, Anhui 230026, China}

\author{Liang-Bi Wu}
\email{wulb@ucas.ac.cn}
\affiliation{School of Fundamental Physics and Mathematical Sciences, Hangzhou Institute for Advanced Study, UCAS, Hangzhou, Zhejiang 310024, China}
 
\author{Yu-Sen Zhou}
\email{zhou\_ys@mail.ustc.edu.cn}
\affiliation{Interdisciplinary Center for Theoretical Study and Department of Modern Physics,\\
University of Science and Technology of China, Hefei, Anhui 230026, China}	

\author{Li-Ming Cao}
\email[corresponding author: ]{caolm@ustc.edu.cn}
\affiliation{Interdisciplinary Center for Theoretical Study and Department of Modern Physics,\\
University of Science and Technology of China, Hefei, Anhui 230026, China}
\affiliation{Peng Huanwu Center for Fundamental Theory, Hefei, Anhui 230026, China}

%<<<<<<<<<<<<< DATE >>>>>>>>>>>>>>>%
\date{\today}
	
%======================================%
%<<<<<<<<<<<<< ABSTRACT >>>>>>>>>>>>>>>%
%======================================%

\begin{abstract}
We investigate the gravitational quasinormal mode spectrum of Kerr--de Sitter black holes with a partially reflective inner boundary near the event horizon, by reducing the angular and radial Teukolsky equations to Heun equations. We vary different parameters and study the $s=-2$, $\ell=m=2$ modes numerically. As the reflectivity $\mathcal{R}_0$ increases, QNMs develop echo branches approaching the real axis with approximately regular spacing, while others approach $\operatorname{Re}\omega=m\Omega(r_+)$. As the rotation $a$ increases, some QNMs approach the real axis or retain finite damping, while others exhibit approximately horizontal shifts with real parts tracking $m\Omega(r_+)$. Two mechanisms driving the instabilities are identified in complementary regimes, namely superradiant amplification and energy injection from the reflective wall. We derive the ratio between Teukolsky amplitude and energy reflectivities, finding that the latter can diverge at $m\Omega(r_+)$ even when the former is small, which accounts for an active wall in this regime. We also find that formulating the reflective boundary condition in terms of asymptotic amplitudes gives rise to resonances near each pole of the ingoing connection coefficient, extending to arbitrarily large positive imaginary parts. An exact finite radius Robin condition imposed on the Sasaki--Nakamura variable removes these pole-born frequencies. These results show how the inner boundary prescription controls the spectrum structure and stability of semi-open Kerr--de Sitter systems.
\end{abstract}

\maketitle
	
%======================================%
%<<<<<<<<<<< Introduction >>>>>>>>>>>>>%
%======================================%

\section{Introduction}\label{sec:introduction}
Quasinormal modes (QNMs) describe the characteristic response of perturbed black holes. They are determined by boundary conditions that select a discrete set of complex frequencies. The real part specifies the oscillation frequency, while the imaginary part controls exponential decay or growth for the time dependence $\mathrm{e}^{-\mathrm{i}\omega t}$. Their dependence on the background geometry makes QNMs central to black-hole spectroscopy and provides a probe of strong-field gravity~\cite{Kokkotas:1999bd,Berti:2009kk,Konoplya:2011qq,Berti:2025hly}. Computing them as solutions of a non-self-adjoint boundary-value problem has a long history~\cite{Leaver:1985ax}.

An important feature of QNM spectra is their sensitivity to perturbations. As shown in early studies by Nollert and Price~\cite{Nollert:1996rf,Nollert:1998ys}, small modifications of the scattering problem may produce large shifts in highly damped resonances. Such behavior is naturally related to the non-Hermitian character of the spectral problem~\cite{Ashida:2020dkc,Konoplya:2022pbc,Jaramillo:2020tuu}. Pseudospectrum analysis provides a framework for quantifying sensitivity to perturbations bounded in a chosen operator norm~\cite{trefethen2020spectra,Jaramillo:2021tmt,Destounis:2023ruj,Boyanov:2024fgc}. This approach has been applied to a range of black-hole backgrounds and related wave equations~\cite{Destounis:2021lum,Cao:2024oud,Cao:2024sot,Boyanov:2023qqf,Cownden:2023dam,Destounis:2023nmb,Chen:2024mon,Cai:2025irl,Cao:2025qws,Zhou:2025xdo,Wu:2026xeg,Zhang:2026cjq}. A pseudospectrum characterizes possible spectral displacements under a prescribed class and norm of perturbations, but it does not by itself identify the displacement produced by a particular physical modification. This motivates studying structured perturbations, such as localized changes in the effective potential~\cite{Qian:2020cnz,Daghigh:2020jyk,Liu:2021aqh,Cheung:2021bol,Berti:2022xfj,Courty:2023rxk,Yang:2024vor,Cardoso:2024mrw,Ianniccari:2024ysv,MalatoCorrea:2025iuc}. Recent investigations have also considered perturbations of the background geometry or of the field equations, including scenarios beyond vacuum general relativity~\cite{Shen:2025nsl,Wang:2025mxe,Hu:2025efp,Mai:2025cva}. 

Another class of modifications concerns the inner boundary condition. In the conventional black-hole problem, the event horizon is purely absorbing. By contrast, certain effective wave models of exotic compact objects (ECOs) replace it with a partially reflecting surface. Near-horizon reflection can produce gravitational-wave echoes through repeated scattering between the surface and the exterior potential barrier~\cite{Cardoso:2016rao,Mark:2017dnq,Bueno:2017hyj,Cardoso:2019apo}. Several models for the reflectivity have been proposed, including the wormhole model~\cite{Mark:2017dnq,Bueno:2017hyj}, the Boltzmann model~\cite{Oshita:2019sat,Wang:2019rcf}, the area quantization model~\cite{Deppe:2024fdo}, the perfect-fluid star, gravastar, membrane paradigm models~\cite{Chakraborty:2023zed}. Reflective boundaries can also cause spectral instability. In rotating configurations, the role of ergoregion instabilities and boundary absorption has been explored in Refs. \cite{Maggio:2017ivp,Maggio:2019qhr}. More generally, the spectral response to modified boundary conditions need not be small even when the modification is weak~\cite{Oshita:2025ibu,Solidoro:2024yxi,Destounis:2025dck,Wu:2025wbp}.

In this work we extend the semi-open boundary-value problem to Kerr--de Sitter (KdS) black holes, which are stationary rotating solutions with positive cosmological constant. The presence of both a reflective wall near the event horizon and a cosmological horizon distinguishes this setting from the asymptotically flat Kerr problem. The perturbation equations can be transformed into Heun equations~\cite{Suzuki:1998vy}. This structure has already proved useful for the computation of KdS QNMs and angular eigenvalues~\cite{Novaes:2019ynt,Hatsuda:2020sbn}. We exploit local Heun solutions and their connection coefficients to formulate the QNM condition for an effective reflecting wall near the event horizon, while retaining the outgoing condition at the cosmological horizon.

Our main focus is on gravitational perturbations with a reflective wall placed near the event horizon in KdS spacetime. We track the QNMs as the reflectivity or black-hole spin changes and identify distinct spectral branches. The reflectivity defined as a ratio of Teukolsky amplitudes should not be identified directly with an energy-flux reflectivity~\cite{Nakano:2017howls,Chen:2021tfc}. Our real-frequency crossing analysis identifies two mechanisms for echo-mode growth, including effective boundary gain in the non-superradiant sector and superradiant feedback in rotating backgrounds. For the asymptotic-amplitude boundary prescription used in the spectral calculations, we also identify pole-born branches associated with simple poles of the ingoing connection coefficient. This feature depends on the boundary prescription. For comparison, we formulate a finite-radius Robin condition on the SN variable, which is more physically motivated. Under this condition, the resonance equation is determined by a finite-radius boundary functional, and a pole of the Teukolsky coefficient $B_{\text{in}}$ alone does not imply a pole-born resonance. 

The rest of this paper is organized as follows. Section~\ref{sec:background} introduces the KdS geometry, the Teukolsky equation, and its reduction to Heun form. Section~\ref{sec:solutions} constructs the local Heun solutions and the semi-open QNM boundary condition. Section~\ref{sec:qnm_spectra} presents the spectral results, followed by the conclusions in Section~\ref{sec:conclusions}. Appendices~\ref{app:flux} and \ref{app:sn} discuss the Teukolsky--Starobinsky current normalization and an SN amplitude transformation, respectively.

\section{Kerr--de Sitter background and Heun formulation}\label{sec:background}
The Kerr--de Sitter (KdS) line element in the coordinates and conventions used here is
\begin{eqnarray}\label{line_element}
    \mathrm{d}s^2=-\frac{\Delta_{\text{r}}}{\Xi^2 \rho^2}(\mathrm{d}t-a\sin^2\theta \mathrm{d}\varphi)^2+\frac{\Delta_\theta \sin^2\theta}{\Xi^2\rho^2}\Big[a\mathrm{d}t-(r^2 + a^2)\mathrm{d}\varphi\Big]^2+\rho^2 \Big(\frac{\mathrm{d}r^2}{\Delta_{\text{r}}} + \frac{\mathrm{d}\theta^2}{\Delta_\theta}\Big)\, ,
\end{eqnarray}
where
\begin{eqnarray}
    \Delta_{\text{r}} &=& (r^2 + a^2)\Big(1-\frac{\alpha_0}{a^2} r^2\Big) - 2Mr =-\frac{\alpha_0}{a^2}(r - r_+)(r - r_{-})(r - r_{+}^{\prime})(r - r_{-}^{\prime})\, , \nonumber\\
    \Delta_\theta &=& 1 + \alpha_0 \cos^2\theta\, , \nonumber\\
    \rho^2 &=& r^2 + a^2 \cos^2\theta\, , \nonumber\\
    \alpha_0 &=& \frac{\Lambda a^2}{3}\, ,\qquad \Xi=1+\alpha_0\, .
\end{eqnarray}
Here $M$ and $a$ denote the mass and rotation parameters, respectively, and $\Lambda>0$ is the cosmological constant. In the parameter range considered below, $\Delta_{\text{r}} = 0$ has four real roots, $r_{\pm}$, $r^{\prime}_{\pm}$, where $r_{\pm} \rightarrow M \pm \sqrt{M^2 - a^2}$ and $r'_{\pm} \rightarrow \pm a/\sqrt{\alpha_0}$ as $\alpha_0 \rightarrow 0$ or $\Lambda \rightarrow 0$. The coordinate singularities $r = r_{+}$ and $r = r_{-}$ correspond to the outer and inner horizons of the black hole, respectively. The other positive root $r = r^{\prime}_{+}$ represents the cosmological horizon, while $r = r'_{-}$ is a negative root outside the physical region. We define the surface gravities and angular velocities at these horizons by
\begin{eqnarray}
    \kappa(r_i) &=& \frac{1}{2\Xi(r_i^2 + a^2)} \frac{\mathrm{d}\Delta_{\text{r}}(r_i)}{\mathrm{d}r}\, , \\
    \Omega(r_i) &=& \frac{a}{r_i^2 + a^2}\, ,
\end{eqnarray}
where $r_i$ represents $r_{\pm},\,r^{\prime}_{\pm}$. In particular, we write $\kappa_+\equiv\kappa(r_+)$ for the event-horizon surface gravity. With the above signed definition, the surface gravity of the cosmological horizon is negative. In the vacuum region, the spin-$s$ Teukolsky master equation on the Kerr--de Sitter black hole takes the form~\cite{Teukolsky:1973,Khanal:1983,Chambers:1994}:
\begin{eqnarray}\label{eq:kdS_teuk_master}
    0&=&\Bigg\{\Delta_{\text{r}}^{-s}\partial_r\big(\Delta_{\text{r}}^{s+1}\partial_r\big)+\partial_x\Big(\Delta_\theta(1-x^2)\partial_x\Big)+\Xi^2\Big[\frac{a^2(1-x^2)}{\Delta_\theta}-\frac{(r^2+a^2)^2}{\Delta_{\text{r}}}\Big]\partial_t^2+2a\Xi^2\Big(\frac{1}{\Delta_\theta}-\frac{r^2+a^2}{\Delta_{\text{r}}}\Big)\partial_t\partial_\varphi\nonumber\\
    &&+\Xi^2\Big[\frac{1}{\Delta_\theta(1-x^2)}-\frac{a^2}{\Delta_{\text{r}}}\Big]\partial_\varphi^2+s\Xi\Big[\frac{(r^2+a^2)\Delta_{\text{r}}'}{\Delta_{\text{r}}}4r-\frac{2\mathrm{i}a\Xi x}{\Delta_\theta}\Big]\partial_t+s\Xi\Big[\frac{a\Delta_{\text{r}}'}{\Delta_{\text{r}}}+\frac{2\mathrm{i}\Xi x}{\Delta_\theta(1-x^2)}-\frac{4\mathrm{i}\alpha_0 x}{\Delta_\theta}\Big]\partial_\varphi\nonumber\\
    &&+s(1-\alpha_0)-2\alpha_0x^2-\frac{\Xi^2s^2x^2}{\Delta_\theta(1-x^2)}\Bigg\}\Psi_s\, ,
\end{eqnarray}
where $x=\cos\theta$. The prime on $\Delta_{\text{r}}$ denotes differentiation with respect to $r$. Using the mode decomposition
\begin{eqnarray}\label{eq:teuk_mode_ansatz}
    \Psi_s(t,r,x,\varphi) =\mathrm{e}^{-\mathrm{i}\omega t+\mathrm{i}m\varphi}S(x)R(r)\, .
\end{eqnarray}
Eq. \eqref{eq:kdS_teuk_master} separates into the angular and radial Teukolsky equations given below.

\subsection{Angular equation}\label{subsec:angular_equation}
The angular Teukolsky equation reads
\begin{equation}\label{eq:angular_teuk}
    \begin{aligned}
    0={}&\bigg[\frac{\mathrm{d}}{\mathrm{d}x}\Big((1+\alpha_0 x^2)(1-x^2)\frac{\mathrm{d}}{\mathrm{d}x}\Big)+\lambda-s(1-\alpha_0)-2\alpha_0 x^2\\
    &\quad+\frac{\Xi^2}{1+\alpha_0 x^2}\Big(c^2x^2-2csx-c^2+2cm+\frac{4\alpha_0}{\Xi}smx-\frac{(m+sx)^2}{1-x^2}\Big)\bigg]S(x)\, .
    \end{aligned}
\end{equation}
where $m$ is the azimuthal number, $c=a\omega$, and $\lambda$ is the separation constant. This equation has regular singular points at $x=\pm1$ and $x=\pm\mathrm{i}/\sqrt{\alpha_0}$, while $x=\infty$ is a removable singularity. We introduce the M\"{o}bius transformation
\begin{eqnarray}\label{eq:mobius_ang}
    z=\frac{(1-\mathrm{i}/\sqrt{\alpha_0})(x+1)}{2(x-\mathrm{i}/\sqrt{\alpha_0})}\, ,
\end{eqnarray}
which maps the singular points as
\begin{eqnarray}
    x=-1 \mapsto z=0\, ,\qquad x=1 \mapsto z=1\, ,\qquad x=-\mathrm{i}/\sqrt{\alpha_0} \mapsto z=z_{\text{a}}\, ,\qquad x=\mathrm{i}/\sqrt{\alpha_0} \mapsto z=\infty\,,\qquad \infty \mapsto z_{\text{a}\infty}\, ,
\end{eqnarray}
with
\begin{eqnarray}
    z_{\text{a}} \equiv -\frac{(1-\mathrm{i}/\sqrt{\alpha_0})^2}{4\mathrm{i}/\sqrt{\alpha_0}}\, ,\qquad z_{\text{a}\infty} \equiv \frac{1-\mathrm{i}/\sqrt{\alpha_0}}{2}\, .
\end{eqnarray}
We then perform the gauge transformation
\begin{eqnarray}\label{eq:S_transform}
    S(x)=z^{A_1}(z-1)^{A_2}(z-z_{\text{a}})^{A_3}(z-z_{\text{a}\infty})\, y_{\text{a}}(z)\, ,
\end{eqnarray}
where the exponents are defined as
\begin{eqnarray}\label{eq:A_def}
    A_1=\frac{m-s}{2}\, ,\qquad A_2=-\frac{m+s}{2}\, ,\qquad A_3=\frac{\mathrm{i}}{2}\Big(\frac{\Xi}{\sqrt{\alpha_0}} c - m\sqrt{\alpha_0} - \mathrm{i} s\Big)\, .
\end{eqnarray}

After these steps, the function \(y_{\text{a}}(z)\) satisfies the standard Heun's equation
\begin{eqnarray}\label{eq:heun_angular_std}
    y_{\text{a}}^{\prime\prime}(z)+\left(\frac{\gamma_{\text{a}}}{z}+\frac{\delta_{\text{a}}}{z-1}+\frac{\epsilon_{\text{a}}}{z-z_{\text{a}}}\right)y_{\text{a}}^{\prime}(z)+\frac{\alpha_{\text{a}}\beta_{\text{a}} z - q_{\text{a}}}{z(z-1)(z-z_{\text{a}})}y_{\text{a}}(z)=0\, ,
\end{eqnarray}
with
\begin{eqnarray}
    \alpha_{\text{a}}=1\, ,\quad \beta_{\text{a}} = 1-s-\mathrm{i} m\sqrt{\alpha_0}+ \mathrm{i} c\left(\sqrt{\alpha_0}+\frac{1}{\sqrt{\alpha_0}}\right)\, ,\quad  \gamma_{\text{a}} = 2A_1+1\, ,\quad  \delta_{\text{a}} = 2A_2+1\, ,\quad  \epsilon_{\text{a}} = 2A_3+1\, ,
\end{eqnarray}
and
\begin{eqnarray}\label{eq:u_def}
    q_{\text{a}} = \frac{\mathrm{i}\lambda}{4\sqrt{\alpha_0}}+\frac12 + A_1 + \Big(m+\frac12\Big)(A_3-A_3^{\star})\, ,
\end{eqnarray}
where \(A_3^{\star}\) is obtained from \(A_3\) by replacing \(\mathrm{i}\) with \(-\mathrm{i}\). The angular Heun parametrization above is understood for nonzero $a$. In the nonrotating limit $a=0$, we directly use the regular spherical limit $\lambda=\ell(\ell+1)-s(s-1)$, rather than solving the Heun equation. 

\subsection{Radial equation}\label{subsec:radial_equation}
The radial Teukolsky equation is
\begin{eqnarray}\label{eq:radial_teuk}
    \bigg[ \Delta_{\text{r}}^{-s} \frac{\mathrm{d}}{\mathrm{d}r} \Big(\Delta_{\text{r}}^{s+1} \frac{\mathrm{d}}{\mathrm{d}r} \Big)
    &+&\frac{\Xi^2 K^2 - \mathrm{i} s \Xi K \Delta_{\text{r}}^{\prime}}{\Delta_{\text{r}}} + 4 \mathrm{i} s \Xi \omega r\nonumber\\
    &-& \frac{2\alpha_0}{a^2} (s+1)(2s+1) r^2 + 2s(1-\alpha_0) - \lambda \bigg] R(r) = 0\, ,
\end{eqnarray}
where \(K(r) = \omega (r^2 + a^2) - a m\).

The point \(r=\infty\) is again a removable singular point and an appropriate transformation yields the standard Heun's equation. We introduce the Möbius transformation
\begin{eqnarray}\label{eq:mobius_rad}
    z = \frac{(r^{\prime}_+ - r_-)(r - r_+)}{(r^{\prime}_+ - r_+)(r - r_-)}\, ,
\end{eqnarray}
which maps the singularities as
\begin{eqnarray}
    r_+ \mapsto 0\, ,\qquad r'_+ \mapsto 1\, ,\qquad r_- \mapsto \infty\, ,\qquad r'_- \mapsto z_{\text{r}}\, ,\qquad \infty \mapsto z_{\text{r}\infty}\, ,
\end{eqnarray}
where
\begin{eqnarray}
    z_{\text{r}} \equiv \frac{(r'_+ - r_-)(r'_- - r_+)}{(r'_+ - r_+)(r'_{-} - r_-)}\, ,\qquad z_{\text{r}\infty} \equiv \frac{r'_+ - r_-}{r'_+ - r_+}\, .
\end{eqnarray}

We then perform the transformation
\begin{eqnarray}\label{eq:R_transform}
    R(r) = z^{B_1} (z-1)^{B_2} (z - z_{\text{r}})^{B_3} (z - z_{\text{r}\infty})^{2s+1} y_{\text{r}}(z)\, ,
\end{eqnarray}
with
\begin{eqnarray}
    B_1 = \frac{\mathrm{i}\Xi K(r_+)}{\Delta_{\text{r}}'(r_+)}\, ,\qquad B_2 = \frac{\mathrm{i}\Xi K(r'_+)}{\Delta_{\text{r}}'(r'_+)}\, ,\qquad B_3 = \frac{\mathrm{i}\Xi K(r'_-)}{\Delta_{\text{r}}'(r'_{-})}\, .
\end{eqnarray}
The function \(y_{\text{r}}(z)\) satisfies the standard Heun's equation
\begin{eqnarray}\label{eq:heun_radial_std}
    y_{\text{r}}''(z)+\Big(\frac{\gamma_{\text{r}}}{z}+\frac{\delta_{\text{r}}}{z-1}+\frac{\epsilon_{\text{r}}}{z-z_{\text{r}}}\Big)y_{\text{r}}'(z)+\frac{\alpha_{\text{r}}\beta_{\text{r}} z - q_{\text{r}}}{z(z-1)(z-z_{\text{r}})}y_{\text{r}}(z)=0\, ,
\end{eqnarray}
with
\begin{eqnarray}\label{heuncoef}
    \alpha_{\text{r}}=2s+1\, ,\quad \beta_{\text{r}} = s+1 - \frac{2\mathrm{i} \Xi K(r_-)}{\Delta_{\text{r}}'(r_-)}\, ,\quad \gamma_{\text{r}} = 2B_1 + s + 1\, ,\quad \delta_{\text{r}} = 2B_2 + s + 1\, ,\quad \epsilon_{\text{r}} = 2B_3 + s + 1\, ,
\end{eqnarray}
and
\begin{eqnarray}
    q_{\text{r}} &=& -\frac{(1+s)(1+2s) r_-'}{r_- - r_-'} 
    - \frac{3\lambda - 2s(3 - a^2\Lambda) + \Lambda (1+s)(1+2s) r_+ (r_+ + r_+')}{\Lambda (r_- - r_-')(r_+ - r_+')}\nonumber \\
    &&+ \frac{2\mathrm{i} (1+2s)(3 + a^2\Lambda)(r_+ r_- \omega + a^2 \omega - a m)}{\Lambda (r_- - r_-')(r_- - r_+)(r_+ - r_+')}\, ,
    \label{eq:v_def}
\end{eqnarray}
Thus, both the angular and radial problems are reduced to the same canonical Heun's equation of the form
\begin{eqnarray}\label{eq:canonical_heun}
    y''(z)+\Big(\frac{\gamma}{z}+\frac{\delta}{z-1}+\frac{\epsilon}{z-z_\star}\Big)y'(z)+\frac{\alpha\beta z - q}{z(z-1)(z-z_\star)}y(z)=0\, ,
\end{eqnarray}
whose local solutions will be introduced in the next section.

%======================================%
%<<<<<<< Heun solutions and boundary conditions >>>>>>>%
%======================================%
\section{Heun function solutions and boundary conditions}\label{sec:solutions}
In this section, we construct the local solutions of the Heun's equation near the singular points $z=0$ and $z=1$, define the connection coefficients, and then derive the angular eigenvalue condition and the radial QNM condition for the semi-open system.

\subsection{Local solutions and connection coefficients}\label{subsec:local_solutions}
For a generic Heun's equation of the form \eqref{eq:canonical_heun}, the two linearly independent Frobenius solutions near $z=0$ are
\begin{eqnarray}\label{eq:y02_gen}
    y_{01}(z) &=& \mathrm{H}\ell(z_\star,q;\alpha,\beta,\gamma,\delta;z)\, ,\nonumber\\
    y_{02}(z) &=&z^{1-\gamma}\,\mathrm{H}\ell(z_\star,(z_\star\delta+\epsilon)(1-\gamma)+q;\alpha+1-\gamma,\beta+1-\gamma,2-\gamma,\delta;z)\, .
\end{eqnarray}
Similarly, near $z=1$, the two solutions are
\begin{eqnarray} \label{eq:y11_gen}
    y_{11}(z) &=& \mathrm{H}\ell(1-z_\star,\alpha\beta-q;\alpha,\beta,\delta,\gamma;1-z)\, ,\nonumber\\
    y_{12}(z) &=& (1-z)^{1-\delta}\,\mathrm{H}\ell(1-z_\star,((1-z_\star)\gamma+\epsilon)(1-\delta)+\alpha\beta-q;\alpha+1-\delta,\beta+1-\delta,2-\delta,\gamma;1-z)\, .
\end{eqnarray}
We denote the angular solutions by \(y_{\text{a}01}, y_{\text{a}02}, y_{\text{a}11}, y_{\text{a}12}\) and the corresponding radial solutions by \(y_{\text{r}01}, y_{\text{r}02}, y_{\text{r}11}, y_{\text{r}12}\). The connection coefficients between the two bases are defined by
\begin{eqnarray}\label{eq:connections_C}
    y_{11}(z)&=&C_{11}\,y_{01}(z)+C_{12}\,y_{02}(z)\, ,\\
    y_{12}(z)&=&C_{21}\,y_{01}(z)+C_{22}\,y_{02}(z)\, ,
\end{eqnarray}
and inversely,
\begin{eqnarray}\label{eq:connections_D}
    y_{01}(z)&=&D_{11}\,y_{11}(z)+D_{12}\,y_{12}(z)\, ,\\
    y_{02}(z)&=&D_{21}\,y_{11}(z)+D_{22}\,y_{12}(z)\, ,
\end{eqnarray}
where the coefficients are expressed in terms of Wronskians. Defining $W[u,v]\equiv uv'-u'v$, where the prime denotes differentiation with respect to $z$, we obtain
\begin{eqnarray}\label{eq:connection_C_wronskian}
    C_{11}=\frac{W[y_{11},y_{02}]}{W[y_{01},y_{02}]}\, ,\quad C_{12}=\frac{W[y_{11},y_{01}]}{W[y_{02},y_{01}]}\, ,\quad C_{21}=\frac{W[y_{12},y_{02}]}{W[y_{01},y_{02}]}\, ,\quad C_{22}=\frac{W[y_{12},y_{01}]}{W[y_{02},y_{01}]}\, .
\end{eqnarray}
Similarly, the coefficients of the inverse connection relations are
\begin{eqnarray}\label{eq:connection_D_wronskian}
    D_{11}=\frac{W[y_{01},y_{12}]}{W[y_{11},y_{12}]}\, ,\quad D_{12}=\frac{W[y_{01},y_{11}]}{W[y_{12},y_{11}]}\, ,\quad D_{21}=\frac{W[y_{02},y_{12}]}{W[y_{11},y_{12}]}\, ,\quad D_{22}=\frac{W[y_{02},y_{11}]}{W[y_{12},y_{11}]}\, .
\end{eqnarray}
In particular, the matrices $\mathbf{C}=(C_{ij})$ and $\mathbf{D}=(D_{ij})$ satisfy $\mathbf{D}=\mathbf{C}^{-1}$.

The ratios of Wronskians are independent of $z$, provided all solutions are analytically continued consistently and the denominators are nonzero. Thus they may be evaluated at any regular point in a common analytic domain, conveniently at $z=1/2$ when the local series converge there. The corresponding angular and radial connection coefficients are denoted by $C_{\text{a}\,ij},D_{\text{a}\,ij}$ and $C_{\text{r}\,ij},D_{\text{r}\,ij}$, respectively.

\subsection{Angular eigenvalue condition}\label{subsec:angular_eigenvalue}
Regularity of the angular function $S(x)$ is required at both boundaries $x=\pm1$. Near $z=0$ ($x=-1$), the transformation~\eqref{eq:S_transform} gives
\begin{eqnarray}
    S(x) \sim z^{A_1} y_{\text{a}}(z)\, .
\end{eqnarray}
The two Frobenius solutions at \(z=0\) have exponents \(0\) and \(1-\gamma_{\text{a}} = -2A_1\). Therefore, if \(A_1\ge 0\), the solution \(y_{\text{a}01}\) (with exponent \(0\)) is regular, while if \(A_1<0\), we must choose \(y_{\text{a}02}\) (with exponent \(-2A_1>0\)) to cancel the divergence. Similarly, near \(z=1\) (\(x=1\)), the exponents are \(0\) and \(1-\delta_{\text{a}} = -2A_2\). If \(A_2\ge 0\), choose \(y_{\text{a}11}\), and if \(A_2<0\), choose \(y_{\text{a}12}\). Thus, depending on the signs of \(A_1\) and \(A_2\), we select a pair \((y_{0}, y_{1})\) as summarised in Table~\ref{tab:angular_choices}. For the global solution to be regular at both endpoints, the two selected local solutions must be linearly dependent, which is equivalent to the condition
\begin{eqnarray}\label{eq:wronskian_condition}
    W[y_{0}, y_{1}] = 0\, .
\end{eqnarray}
This complex equation determines the angular eigenvalue $\lambda$ as a function of \(\omega\) and the background parameters. We denote the implicit solution by \(\lambda = \lambda_{s,\ell,m}(\omega)\).

\begin{table}[t]
\centering
\begin{ruledtabular}
\begin{tabular}{cccc}
Case & Sign of \(A_1\) & Sign of \(A_2\) & Chosen pair \((y_{0},y_{1})\) \\ \hline
1 & \(A_1\ge 0\) & \(A_2\ge 0\) & \((y_{\text{a}01},\,y_{\text{a}11})\) \\
2 & \(A_1\ge 0\) & \(A_2<0\)   & \((y_{\text{a}01},\,y_{\text{a}12})\) \\
3 & \(A_1<0\)  & \(A_2\ge 0\) & \((y_{\text{a}02},\,y_{\text{a}11})\) \\
4 & \(A_1<0\)  & \(A_2<0\)   & \((y_{\text{a}02},\,y_{\text{a}12})\) \\
\end{tabular}
\end{ruledtabular}
\caption{Selections of local angular solutions according to the signs of $A_1=(m-s)/2$ and $A_2=-(m+s)/2$.}
\label{tab:angular_choices}
\end{table}

\subsection{Radial solutions and QNM condition}\label{subsec:radial_qnm_condition}
The radial Teukolsky equation \eqref{eq:radial_teuk} is reduced to a Heun problem with local solutions $y_{\text{r}01}, y_{\text{r}02}, y_{\text{r}11}, y_{\text{r}12}$. The corresponding radial functions follow by restoring the prefactor
\begin{eqnarray}
    R_{01}(r) &=& z^{B_1} (z-1)^{B_2} (z-z_{\text{r}})^{B_3} (z - z_\infty)^{2s+1} y_{\text{r}01}(z)\sim (r - r_+)^{\mathrm{i}\omega_{\text{h}}/(2\kappa(r_+))}\, ,\\
    R_{02}(r) &=& z^{B_1} (z-1)^{B_2}(z-z_{\text{r}})^{B_3} (z - z_\infty)^{2s+1} y_{\text{r}02}(z)\sim (r - r_+)^{-s-\mathrm{i}\omega_{\text{h}}/(2\kappa(r_+))}\, ,\\
    R_{11}(r) &=& z^{B_1} (z-1)^{B_2} (z-z_{\text{r}})^{B_3} (z - z_\infty)^{2s+1} y_{\text{r}11}(z)\sim (r'_+ - r)^{\mathrm{i}\omega_{\text{c}}/(2\kappa(r'_+))},\\
    R_{12}(r) &=& z^{B_1} (z-1)^{B_2} (z-z_{\text{r}})^{B_3} (z - z_\infty)^{2s+1} y_{\text{r}12}(z)\sim (r'_+ - r)^{-s - \mathrm{i}\omega_{\text{c}}/(2\kappa(r^{\prime}_+))}\, ,
\end{eqnarray}
where
\begin{eqnarray}\label{eq:two_horizon_frequencies}
    \omega_{\text{h}}=\omega-m\Omega(r_+)\, ,\qquad \omega_{\text{c}}=\omega-m\Omega(r^{\prime}_+)\, .
\end{eqnarray}

To match the asymptotic behaviours at the horizons, we define the ``in'', ``out'', ``up'', and ``down'' solutions whose asymptotic behaviors at the event horizon ($r\to r_+$, $r_\star\to-\infty$) and the cosmological horizon ($r\to r'_+$, $r_\star\to+\infty$) are, respectively,
\begin{eqnarray}\label{eq:in_two_horizon_asymptotic}
\Psi_{\text{in}}\sim
\begin{cases}
    \Delta_{\text{r}}^{-s}\mathrm{e}^{-\mathrm{i}\omega_{\text{h}}r_\star}\, , & r\to r_+\, ,\\
    A_{\text{in}}\Delta_{\text{r}}^{-s}\mathrm{e}^{-\mathrm{i}\omega_{\text{c}} r_\star}+A_{\text{out}}\mathrm{e}^{\mathrm{i}\omega_{\text{c}} r_\star}\, , & r\to r'_+\, ,
\end{cases}
\end{eqnarray}
\begin{eqnarray}\label{eq:out_two_horizon_asymptotic}
\Psi_{\text{out}}\sim
\begin{cases}
    \mathrm{e}^{\mathrm{i}\omega_{\text{h}} r_\star}\, , & r\to r_+\, ,\\
    \widetilde{A}_{\text{in}}\Delta_{\text{r}}^{-s}\mathrm{e}^{-\mathrm{i}\omega_{\text{c}} r_\star}+\widetilde{A}_{\text{out}}\mathrm{e}^{\mathrm{i}\omega_{\text{c}} r_\star}\, , & r\to r'_+\, ,
\end{cases}
\end{eqnarray}
\begin{eqnarray}\label{eq:up_two_horizon_asymptotic}
\Psi_{\text{up}}\sim
    \begin{cases}
    B_{\text{in}}\Delta_{\text{r}}^{-s}\mathrm{e}^{-\mathrm{i}\omega_{\text{h}} r_\star}+B_{\text{out}}\mathrm{e}^{\mathrm{i}\omega_{\text{h}} r_\star}\, , & r\to r_+\, ,\\
    \mathrm{e}^{\mathrm{i}\omega_{\text{c}} r_\star}\, , & r\to r'_+\, ,
    \end{cases}
\end{eqnarray}
and
\begin{eqnarray}\label{eq:down_two_horizon_asymptotic}
\Psi_{\text{down}}\sim
    \begin{cases}
        \widetilde B_{\text{in}}\Delta_{\text{r}}^{-s}\mathrm{e}^{-\mathrm{i}\omega_{\text{h}}r_\star}+\widetilde B_{\text{out}}\mathrm{e}^{\mathrm{i}\omega_{\text{h}} r_\star}, & r\to r_+\, ,\\
        \Delta_{\text{r}}^{-s}\mathrm{e}^{-\mathrm{i}\omega_{\text{c}} r_\star}, & r\to r'_+\, .
    \end{cases}
\end{eqnarray}
where the tortoise coordinate $r_\star$ is defined by
\begin{eqnarray}
    \frac{\mathrm{d}r_\star}{\mathrm{d}r}=\Xi\frac{r^2+a^2}{\Delta_{\text{r}}}\, ,
\end{eqnarray}
which integrates to
\begin{eqnarray}
    r_\star = \frac{\ln|r-r_+|}{2\kappa(r_+)} + \frac{\ln|r-r_-|}{2\kappa(r_-)} + \frac{\ln|r-r'_+|}{2\kappa(r'_+)} + \frac{\ln|r-r'_-|}{2\kappa(r'_-)}\, .
\end{eqnarray}
The four solutions \(\Psi_{\text{in}}\),\(\Psi_{\text{out}}\),\(\Psi_{\text{up}}\), and \(\Psi_{\text{down}}\) are proportional to $R_{02}$, $R_{01}$, $R_{11}$, and $R_{12}$, respectively, i.e.,
\begin{eqnarray}
    \Psi_{\text{in}} = \frac{R_{02}(r)}{F(\omega)}\, ,\quad
    \Psi_{\text{out}} = \frac{R_{01}(r)}{G(\omega)}\, ,\quad
    \Psi_{\text{up}} = \frac{R_{11}(r)}{H(\omega)}\, ,\quad  \Psi_{\text{down}} = \frac{R_{12}(r)}{I(\omega)}\, .
\end{eqnarray}
The explicit expressions for the normalization factors $F(\omega),G(\omega),H(\omega),I(\omega)$ are
\begin{eqnarray}
    F(\omega) &=& (-1)^{B_2+B_3+2s+1}(\Lambda/3)^s(r'_+-r_+)^{B_1-B_3-1+P_1}(r'_+-r_-)^{B_3-B_1+s+1}(r_+-r_-)^{B_1+2s+P_2}\nonumber\\
    &&\times(r_+-r'_-)^{B_3+s+P_3}(r_--r'_-)^{-B_3}\, ,\\
    G(\omega) &=& (-1)^{B_2+B_3+2s+1}(r'_+-r_+)^{-B_1-B_3-2s-1-P_1}(r'_+-r_-)^{B_3+B_1+2s+1}(r_+-r_-)^{-B_1-P_2}\nonumber\\
    &&\times(r_+-r'_-)^{B_3-P_3}(r_--r'_-)^{-B_3}\, ,\\
    H(\omega) &=& (-1)^{B_2+B_3+2s+1}(r'_+-r_+)^{-B_2-B_3-2s-1-Q_1}(r'_+-r_-)^{-B_2-Q_2}(r'_+-r'_-)^{B_3-Q_3}\nonumber\\
    &&\times(r_+-r_-)^{B_2+B_3+2s+1}(r_--r'_-)^{-B_3}\, ,\\
    I(\omega) &=& (-1)^{B_2+B_3+2s+1}(\Lambda/3)^s(r'_+-r_+)^{B_2-B_3-1+Q_1}(r'_+-r_-)^{B_2+2s+Q_2}(r'_+-r'_-)^{B_3+s+Q_3}\nonumber\\
    &&\times(r_+-r_-)^{-B_2+B_3+s+1}(r_--r'_-)^{-B_3}\, ,
\end{eqnarray}
where the above auxiliary exponents are
\begin{eqnarray}
    P_1 &=& \frac{\mathrm{i}\omega_\text{h}}{2\kappa(r'_+)}\, ,\qquad P_2 = \frac{\mathrm{i}\omega_\text{h}}{2\kappa(r_-)}\, ,\qquad P_3 = \frac{\mathrm{i}\omega_\text{h}}{2\kappa(r'_-)}\, ,\\
    Q_1 &=& \frac{\mathrm{i}\omega_{\text{c}}}{2\kappa(r_+)}\, ,\qquad Q_2 = \frac{\mathrm{i}\omega_{\text{c}}}{2\kappa(r_-)}\, ,\qquad Q_3 = \frac{\mathrm{i}\omega_{\text{c}}}{2\kappa(r'_-)}\, .
\end{eqnarray}
Using the connection coefficients $D_{\text{r}\,ij}$ and $C_{\text{r}\,ij}$ for the radial sector, the asymptotic coefficients are expressed as
\begin{eqnarray}\label{eq:ABcoeff_full}
    A_{\text{in}}(\omega) &=& D_{\text{r}\,22}\,\frac{I(\omega)}{F(\omega)}\, ,\qquad
    A_{\text{out}}(\omega) = D_{\text{r}\,21}\,\frac{H(\omega)}{F(\omega)}\, ,\\
    B_{\text{in}}(\omega) &=& C_{\text{r}\,12}\,\frac{F(\omega)}{H(\omega)}\, ,\qquad
    B_{\text{out}}(\omega) = C_{\text{r}\,11}\,\frac{G(\omega)}{H(\omega)}\, .
\end{eqnarray}
These expressions are obtained by matching the asymptotic behaviours at $r\to r_+$ and $r\to r^{\prime}_+$, respectively.

We now define a near-horizon reflecting boundary located at $r_{\star\text{w}}$ which is characterized by the solution~\cite{Mark:2017dnq,Chen:2021tfc}
\begin{eqnarray}\label{eq:semi_open_BC_full}
    \Psi_{\text{wall}}(r,\omega)=\Psi_{\text{in}}(r,\omega)+\mathcal R_0 \mathrm{e}^{-2\mathrm{i}\omega_\text{h}r_{\star\text{w}}}\Psi_{\text{out}}(r,\omega)\, ,
\end{eqnarray}
where $\mathcal{R}_0$ is the reflection amplitude and note that $\mathcal R_0$ is dimensionful. The factor $\mathrm{e}^{-2\mathrm{i}\omega_\text{h}r_{\star\text{w}}}$ incorporates the leading round-trip phase from $r_{\star\text{w}}$. We consider a reflective boundary near the event horizon, $r_{\star\text{w}}<0$ and $|r_{\star\text{w}}|\gg1$, for which the boundary condition becomes
\begin{eqnarray}
    \Psi_{\text{wall}}(r_{\star})\sim\Delta_{\text{r}}^{-s}\mathrm{e}^{-\mathrm{i}\omega_\text{h}r_\star}+\mathcal {R}_0\mathrm{e}^{-2\mathrm{i}\omega_\text{h}r_{\star\text{w}}}\mathrm{e}^{\mathrm{i}\omega_{\text{h}}r_\star}\, ,\qquad r_\star\to r_{\star\text{w}}\, .
\end{eqnarray}
At the cosmological horizon, the solution has the form
\begin{eqnarray}
    \Psi_{\text{wall}}\sim S_{\text{in}}(\omega)\Delta_{\text{r}}^{-s}\mathrm{e}^{-\mathrm{i}\omega_{\text{c}}r_\star}+S_{\text{out}}(\omega)\mathrm{e}^{\mathrm{i}\omega_{\text{c}}r_\star}\, ,\qquad r_\star\to+\infty\, .
\end{eqnarray}

The QNM condition $S_{\text{in}}=0$ requires $\Psi_{\text{wall}}$ to be proportional to $\Psi_{\text{up}}$. Using the exact connection relation
\begin{eqnarray}\label{eq:connection}
    \Psi_{\text{up}}=B_{\text{in}}\Psi_{\text{in}}+B_{\text{out}}\Psi_{\text{out}}\, ,
\end{eqnarray}
and comparing the coefficients of the independent near-horizon basis solutions, we obtain the QNM condition
\begin{eqnarray}\label{eq:qnm_condition_raw_full}
    B_{\text{in}}(\omega,\lambda)\, \mathcal{R}_0\, \mathrm{e}^{-2\mathrm{i}\omega_{\text{h}} r_{\star\text{w}}} - B_{\text{out}}(\omega,\lambda) = 0\, .
\end{eqnarray}
We introduce the reflection function
\begin{eqnarray}\label{eq:Rfunc_full}
    \mathcal{R}(\omega,\lambda):=\frac{B_{\text{out}}(\omega,\lambda)}{B_{\text{in}}(\omega,\lambda)}\mathrm{e}^{2\mathrm{i}\omega_{\text{h}} r_{\star\text{w}}}\, ,
\end{eqnarray}
and combine Eq. \eqref{eq:wronskian_condition} with Eq. \eqref{eq:qnm_condition_raw_full}, we obtain
\begin{eqnarray}\label{eq:qnm_condition_full}
    \mathcal{R}(\omega,\lambda_{s,\ell,m}(\omega))=\mathcal{R}_0\, .
\end{eqnarray}
With the angular eigenvalue specified by $\lambda=\lambda_{s,\ell,m}(\omega)$, Eq. \eqref{eq:qnm_condition_full} determines the discrete QNM spectra $\omega_{\ell mn}$ for given background parameters, reflection amplitude $\mathcal{R}_0$, and wall position $r_{\star\text{w}}$. At exactly $\mathcal{R}_0=0$, the original resonance condition \eqref{eq:qnm_condition_raw_full} reduces to $B_{\text{out}}=0$, recovering the standard Kerr--dS black-hole QNM boundary conditions. Note that the limit $\mathcal{R}_0\to0^+$ does not reproduce the case $\mathcal{R}_0=0$ as discussed in Sec. \ref{sec:qnm_spectra}. 

\section{Quasinormal mode spectra}\label{sec:qnm_spectra}
We now investigate how the QNM spectra depend on the black-hole rotation parameter $a$ and the reflection amplitude $\mathcal R_0$. We denote the mode frequencies by $\omega_{\ell mn}$, where $\ell$ labels the angular eigenvalue branch through its nonrotating limit, as specified in Sec. \ref{subsec:angular_equation}, and $n$ is the overtone index. At $\mathcal R_0=0$ and $a=0$, the spectrum is symmetric under reflection across the imaginary-frequency axis. We use ``+'' and ``-'' to distinguish the branches with positive and negative real parts of the frequency, respectively. The overtone index $n$ is assigned in this reference configuration and continued along the chosen path in $(\mathcal R_0,a)$ space. In the presence of exceptional points, this continuation can be path dependent, so the mode labels need not be globally unique. Unless stated otherwise, the numerical calculation uses $M=0.5$, $\Lambda=0.04$, $r_{\star\text{w}}=-100$, $s=-2$, and $\ell=m=2$. The modes were computed with $100$-digit working precision. Convergence was checked by repeating the calculations with $140$- and $180$-digit working precision. For the tested modes, the absolute differences in both $\omega$ and $\lambda$ were below $10^{-100}$.

\begin{figure}[H]
    \centering
    \subfigure[]{\includegraphics[width=0.28\linewidth]{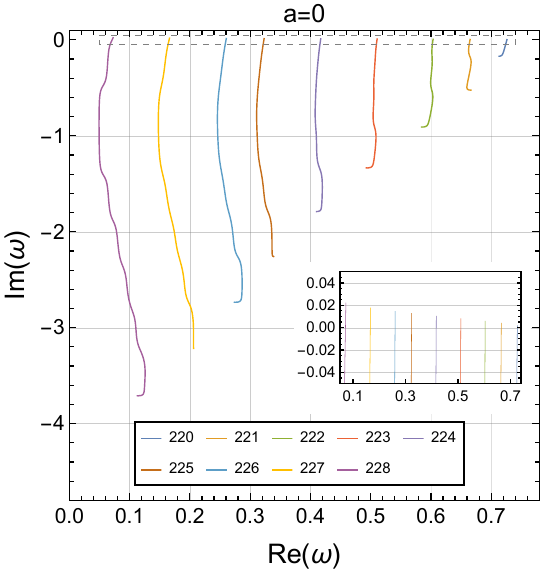}\label{fig:olofR0}}  \hfill
    \subfigure[]{\includegraphics[width=0.28\linewidth]{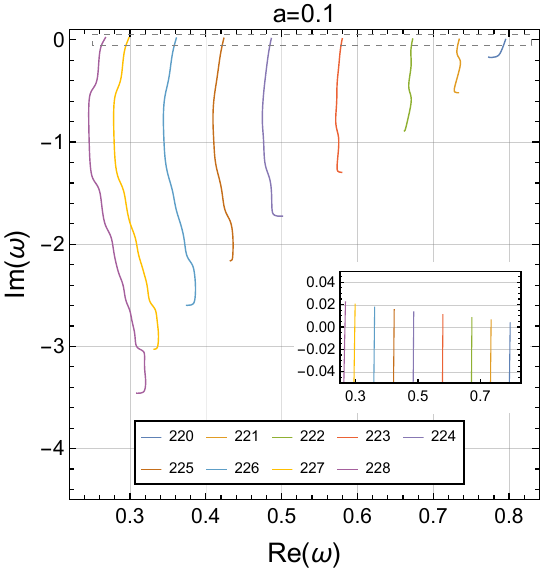}\label{fig:olofR1}} \hfill
    \subfigure[]{\includegraphics[width=0.28\linewidth]{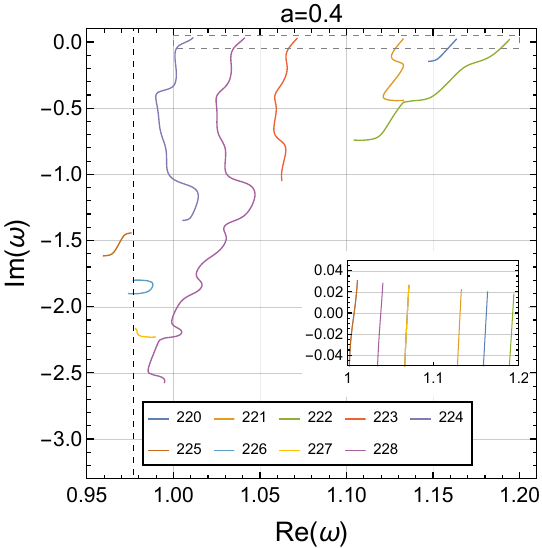}\label{fig:olofR4}}
    
    \caption{``+'' QNMs for fixed $a$ with $\mathcal{R}_0$ varying from $10^{-440}$ to $1$. The dashed vertical line in panel~\ref{fig:olofR4} marks $\mathrm{Re}(\omega)=m\Omega(r_+)$. The enlarged images of the dashed boxes are shown in the insets.}
    \label{fig:olofRp}
\end{figure}

\begin{figure}[H]
    \centering
    \subfigure[]{\includegraphics[width=0.28\linewidth]{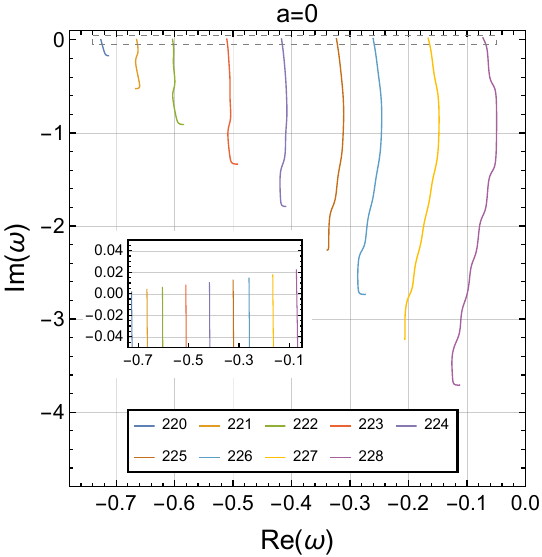}}  \hfill
    \subfigure[]{\includegraphics[width=0.28\linewidth]{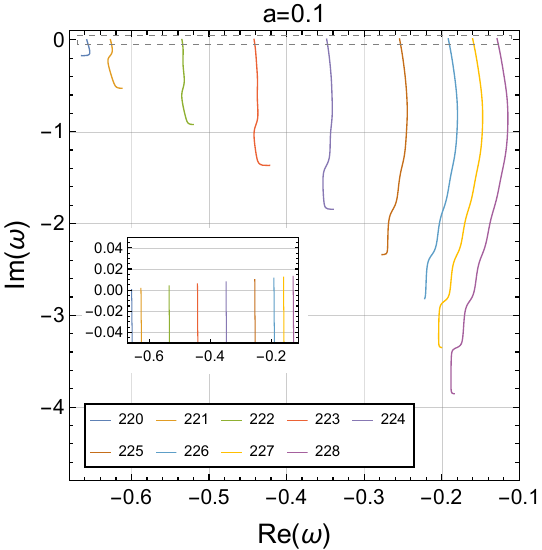}} \hfill
    \subfigure[]{\includegraphics[width=0.28\linewidth]{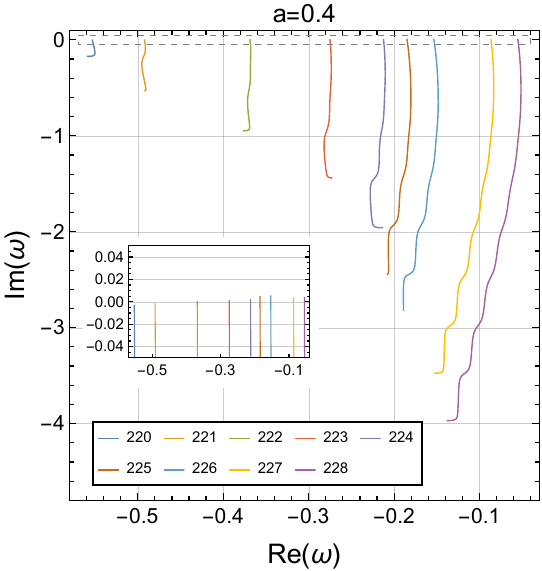}}
    
    \caption{``-'' QNMs for fixed $a$ with $\mathcal{R}_0$ varying from $10^{-440}$ to $1$. The enlarged views of the dashed boxes are shown in the insets.}
    \label{fig:olofRm}
\end{figure}

To isolate the effect of the reflective boundary at a specified spin, we first continue the ordinary black-hole modes from $(\mathcal{R}_0,a)=(0,0)$ to $(0,a)$ and then increase $\mathcal{R}_0$ while holding $a$ fixed. The resulting trajectories for the two frequency branches are shown in Figs. \ref{fig:olofRp} and \ref{fig:olofRm}.

As $\mathcal{R}_0$ increases, the plotted trajectories generally shift toward larger imaginary parts. The trajectories show two broad patterns. Many approach the real-frequency axis at high reflectivity. We refer to these modes, whose eigenfunction are trapped between the reflective boundary and the exterior scattering region, as echo modes. Other trajectories like the modes labeled ``225'' and ``226'' in Fig. \ref{fig:olofR4} approach the vertical line $\operatorname{Re}\omega=m\Omega(r_+)$.

The approximately regular spacing of the echo modes follows from the phase accumulated during propagation to and from the wall. Near the real-frequency axis, the reflection function $\mathcal{R}(\omega,\lambda_{s,\ell,m}(\omega))$ contains the factor $\exp(2\mathrm{i}\omega_{\text{h}} r_{\star\text{w}})$ [see Eq. (\ref{eq:qnm_condition_raw_full})]. If $B_{\text{out}}/B_{\text{in}}$ varies slowly over the relevant frequency interval, the phase of this exponential dominates the resonance condition. For a fixed real, positive $\mathcal{R}_0$, neighboring modes therefore have approximately the same imaginary part and a real-frequency separation $\Delta\operatorname{Re}\omega\simeq\pi/|r_{\star\text{w}}|\simeq0.03$ for the wall position used in the numerical calculation ($r_{\star\text{w}}=-100$). 

\begin{figure}[H]
    \centering
    \subfigure[]{\includegraphics[width=0.28\linewidth]{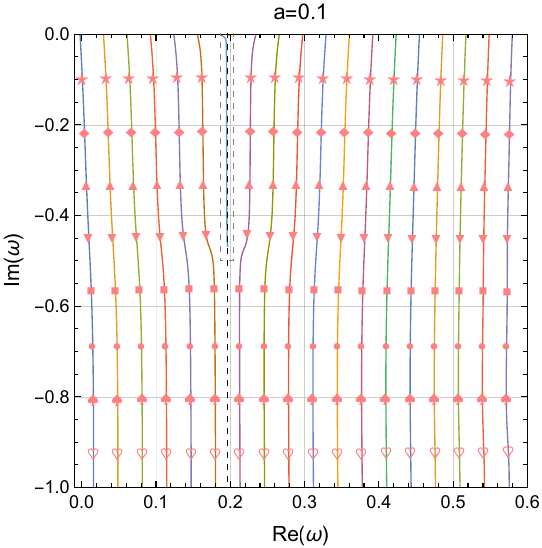}\label{fig:contoura1}}\hspace{1.5cm}
    \subfigure[]{\includegraphics[width=0.28\linewidth]{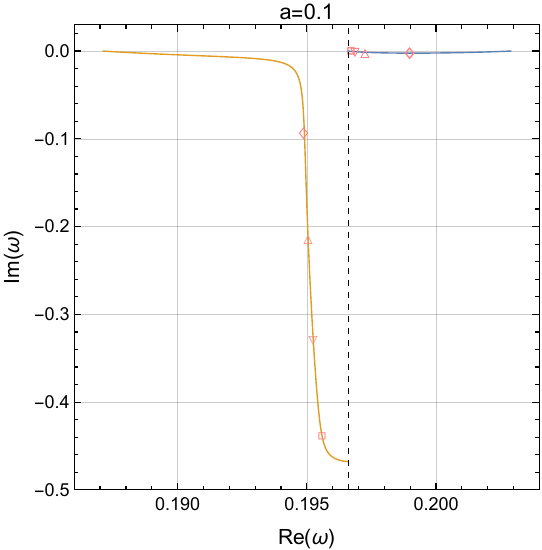}\label{fig:smallcontoura1}}
    
    \caption{The contours on which $\mathcal{R}(\omega,\lambda_{s,\ell,m}(\omega))$ is real and positive for $a=0.1$ are shown in the left panel, while the enlarged view of the dashed boxed area is shown in the right panel. The dashed vertical line is $\mathrm{Re}(\omega)=m\Omega(r_+)$. Markers of different shapes correspond to different values of the reflection amplitude $\mathcal R_0$. In the left panel, $\heartsuit=10^{-80}$, $\spadesuit=10^{-70}$, $\bullet=10^{-60}$, $\blacksquare=10^{-50}$, $\blacktriangledown=10^{-40}$, $\blacktriangle=10^{-30}$, $\blacklozenge=10^{-20}$, $\star=10^{-10}$. In the right panel, for the yellow mode $\square=10^{-40}$, $\triangledown=10^{-30}$, $\triangle=10^{-20}$, $\diamondsuit=10^{-10}$; while for the blue mode $\square=10^{-4}$, $\triangledown=10^{-3.75}$, $\triangle=10^{-3.5}$, $\diamondsuit=10^{-3.25}$.}
    \label{fig:contour}
\end{figure}

The displayed trajectories in Figs. \ref{fig:olofRp} and \ref{fig:olofRm} represent selected modes continued from the Kerr--dS spectrum. Consequently, the spacing between two displayed curves can be a multiple of $\Delta\operatorname{Re}\omega$ if intermediate branches have not been included. Fig. \ref{fig:contour} provides a complementary view in which contours show where $\mathcal{R}(\omega,\lambda_{s,\ell,m}(\omega))$ is real and positive at fixed spin. By Eq. \eqref{eq:qnm_condition_full}, for positive real $\mathcal{R}_0$, these contours contain the resonance trajectories as $\mathcal{R}_0$ is varied. They reveal additional branches that are not present among the selected overtone trajectories in Fig. \ref{fig:olofR1}, and make the approximate spacing more apparent.

Some echo mode trajectories cross the real-frequency axis and enter the upper half-plane as $\mathcal R_0$ increases, as illustrated in the insets of Figs. \ref{fig:olofRp} and \ref{fig:olofRm}. To understand the underlying mechanism, we need to distinguish the prescribed Teukolsky amplitude ratio $\mathcal R_0$ from the corresponding energy flux reflectivity. In general, $|\mathcal R_0|^2$ is not the ratio of reflected to incident energy flux~\cite{Nakano:2017howls,Chen:2021tfc}. We therefore define the ratio of energy fluxes for real frequencies by
\begin{eqnarray}\label{eq:flux_reflectivity}
    \mathcal R_{\text{E}}(\omega)=\frac{\mathcal F_{\text{out}}}{\mathcal F_{\text{in}}}=\eta(\omega,\ell,m,a,\Lambda)|\mathcal R_0|^2\, ,
\end{eqnarray}
where the energy fluxes $\mathcal F_{\text{out}}$ and $\mathcal F_{\text{in}}$ and the ratio $\eta$ follow from the gravitational Teukolsky--Starobinsky (TS) normalization and are defined in Eq. \eqref{eq:app_flux_waves_current} and \eqref{eq:app_flux_eta} of Appendix \ref{app:flux}. Appendix \ref{app:sn} also gives the derivation of $\mathcal R_{\text{E}}(\omega)$ and $\eta_{\text{SN}}$ in Sasaki--Nakamura formalism in Eq. \eqref{eq:app_sn_eta_SN}. 

A wall is called flux-passive at a given real frequency $\omega$ if
\begin{eqnarray}\label{eq:wall_passivity}
    \mathcal R_{\text{E}}(\omega)\leq 1\, ,\qquad    |\mathcal R_0|\leq\eta(\omega)^{-1/2}\, .
\end{eqnarray}
Otherwise, it is active, reflecting more energy flux than it receives and thus supplying energy to the perturbation. Thus $|\mathcal R_0|<1$ does not by itself imply that the boundary is passive. Note that $\eta$ diverges at $\omega=m\Omega(r_+)$, as shown in Appendix \ref{app:flux}. Hence, even for small $\mathcal R_0$, the wall becomes active sufficiently close to $\omega=m\Omega(r_+)$.

\begin{figure}[t]
 \centering

  \includegraphics[width=0.6\linewidth]{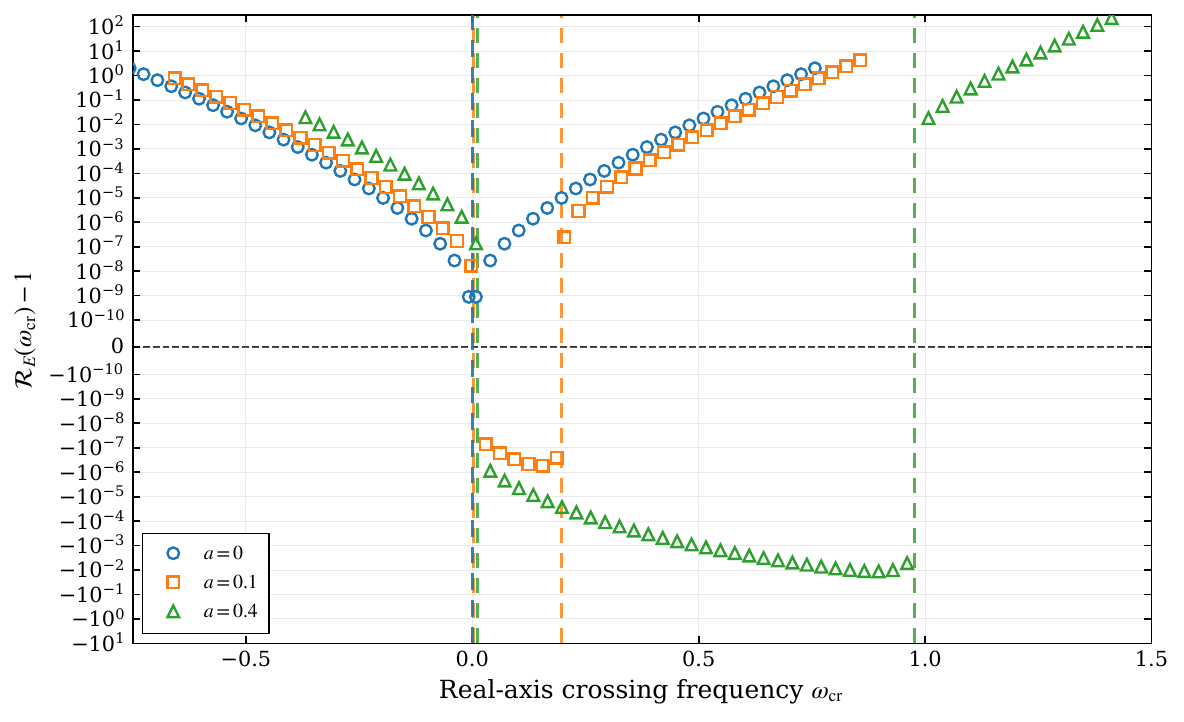}%
 \caption{Real-axis echo mode crossings for $M=0.5$, $\Lambda=0.04$, $s=-2$, $\ell=m=2$, $r_{\star\text{w}}=-100$, and $0<\mathcal R_{0,\text{cr}}\leq1$. Signed flux excess $\mathcal R_{\text{E}}(\omega_{\rm cr})-1$ as a function of the crossing frequency. The dashed vertical lines indicate the two synchronous frequencies $m\Omega(r_+^{\prime})$ and $m\Omega(r_+)$ for each rotating case. The color of the vertical line is the same as the corresponding data points. Note that for $a=0.1$ and $a=0.4$, $m\Omega(r_+^{\prime})$ is smaller than $m\Omega(r_+)$ and lie close to zero, while for $a=0$ they are both equal to zero and the two vertical lines overlap.}
 \label{fig:flux_crossings_pair}
\end{figure}

We can also find directly where echo modes cross the real axis, without following the trajectories in the figures. The threshold frequency $\omega_{\rm cr}\in\mathbb{R}$ must satisfy
\begin{eqnarray}\label{eq:real_axis_crossing_test}
    \operatorname{Im}\mathcal R(\omega_{\rm cr},\lambda_{-2,2,2}(\omega_{\rm cr}))=0\, ,\qquad\mathcal R_{0,\text{cr}}=\operatorname{Re}\mathcal R (\omega_{\rm cr},\lambda_{-2,2,2}(\omega_{\rm cr}))\in(0,1]\, ,
\end{eqnarray}
with $\mathcal R$ defined in Eq. \eqref{eq:Rfunc_full}. Rather than fixing $\mathcal{R}_0$ and searching for the threshold frequency $\omega_{\rm cr}$, here we scan the real frequencies and determine the critical pairs $(\omega_{\rm cr},\mathcal R_{0,\text{cr}})$. Different crossings generally correspond to different constant values of $\mathcal R_0$. For the parameters of Figs. \ref{fig:olofRp} and \ref{fig:olofRm}, we evaluate the scattering coefficients using the angular eigenvalue $\lambda_{-2,2,2}(\omega)$ continued along the $\ell=m=2$ branch at each real frequency. Candidate crossings are identified by solving $\operatorname{Im}\mathcal R(\omega,\lambda_{-2,2,2}(\omega))=0$ along the real-frequency axis. Only pairs satisfies $\mathcal R_{0,\text{cr}} =\operatorname{Re}\mathcal R(\omega_{\rm cr}, \lambda_{-2,2,2}(\omega_{\rm cr}))\in(0,1]$ are retained. Even for $\mathcal R_{0,\text{cr}}<1$, the corresponding $\mathcal R_{\text{E}}$ may exceed unity. Our real-frequency scan over $-3\leq\omega\leq3$ finds crossings summarized in Fig. \ref{fig:flux_crossings_pair}, including those not contained in Figs. \ref{fig:olofRp} and \ref{fig:olofRm}. The presence of two horizons is crucial for interpreting these crossings. In terms of the horizon frequencies introduced in Eq. \eqref{eq:two_horizon_frequencies}, for $a>0$, the superradiant interval is
\begin{eqnarray} \label{eq:superradiant_band}
    m\Omega(r^{\prime}_+)<\omega<m\Omega(r_+)\, ,\qquad \omega_{\text{c}}>0>\omega_{\text{h}}\, .
\end{eqnarray}
For $a=0.1$, the endpoints are approximately $0.0030$ and $0.20$, while for $a=0.4$, they are $0.012$ and $0.98$, respectively. The $a=0$ case contains no such interval and all its crossings satisfy $\mathcal R_{\text{E}}>1$.

All the crossings with $\mathcal R_{\mathrm{E}}<1$ lie inside the superradiant interval. This follows from a flux argument. Let $A_{\rightarrow}$ and $A_{\leftarrow}$ denote the current-normalized amplitudes of waves propagating outward from and inward at the event horizon, respectively. At the cosmological horizon, let $T$ denote the current-normalized amplitudes of waves propagating outward. Their definitions are given in  Appendix \ref{app:flux}. The Wronskian for the real frequency gives
\begin{eqnarray} \label{eq:barrier_current}
    \omega_{\text{h}}\bigl(|A_{\rightarrow}|^2-|A_{\leftarrow}|^2\bigr)=\omega_{\text{c}}|T|^2\, ,
\end{eqnarray}
which is a physical gravitational energy-flux relation. For critical real frequency $\omega_{\rm cr}$, the energy reflectivity is 
\begin{eqnarray}\label{eq:barrier_gain}
    \mathcal R_{\text{E}}(\omega_{\rm cr})=\frac{\mathcal F_{\text{out}}}{\mathcal F_{\text{in}}}=\frac{|A_{\rightarrow}|^2}{|A_{\leftarrow}|^2}=1+\frac{\omega_{\text{c}}}{\omega_{\text{h}}}\frac{|T|^2}{|A_{\leftarrow}|^2}\, .
\end{eqnarray}
For $\omega_{\text{h}}\omega_{\text{c}}>0$ the reflective wall is active, $\mathcal R_{\text{E}}(\omega_{\rm cr})>1$, whereas for $\omega_{\text{h}}\omega_{\text{c}}<0$ the reflective wall is passive, $\mathcal R_{\text{E}}(\omega_{\rm cr})<1$, which is also numerically verified by Fig. \ref{fig:flux_crossings_pair}. This implies that outside the superradiant regime, the instability is driven by energy injected through the reflective wall, whereas inside this regime, it is driven by superradiance. Accordingly, the crossings in the nonrotating case require an active reflective wall. For rotating black holes, by contrast, instabilities caused by superradiance may arise even when the reflective boundary itself absorbs energy. 

To understand the second class of trajectories in Fig. \ref{fig:olofR4}, whose real parts approach $\operatorname{Re}(\omega)=m\Omega(r_+)$ as $\mathcal R_0$ tends to $1$, we examine Eqs. \eqref{eq:Rfunc_full} and \eqref{eq:qnm_condition_full}. For a wall deep in the near-horizon tortoise region, $r_{\star\text{w}}<0$ and $|r_{\star\text{w}}|\gg1$, the factor $\mathrm{e}^{2\mathrm{i}\omega_{\mathrm h}r_{\star\text{w}}}$ is exponentially suppressed at fixed $\operatorname{Im}\omega<0$. Consequently, poles of $B_{\text{out}}/B_{\text{in}}$ can compensate for this suppression, admitting solutions of $\mathcal R(\omega)=\mathcal R_0$ in their vicinity. We therefore examine the poles of the connection coefficients as well as the zeros of their ratio.

The analytic structure of the Heun connection coefficients involves gamma-function factors~\cite{Bonelli:2022ten}. In the present Frobenius normalization, the relevant factors are $\Gamma(2-\gamma_{\text{r}})$ in $C_{\text{r}\,11}$ and $\Gamma(\gamma_{\text{r}})$ in $C_{\text{r}\,12}$, up to other parameter-dependent factors. Together with Eq. \eqref{eq:ABcoeff_full}, they indicate candidate poles of $B_{\rm out}$ at $\gamma_{\text{r}}=2,3,\cdots$ and of $B_{\rm in}$ at $\gamma_{\text{r}}=0,-1,\cdots$. Using Eq. \eqref{heuncoef}, their candidate locations are
\begin{eqnarray}\label{eq:heun_candidate_poles}
    B_{\text{in}}: \quad\omega&=&m\Omega(r_+)+\mathrm{i}\kappa_+(s+k)\, ,\qquad k=1,2,\cdots\, ,\\
    B_{\text{out}}: \quad\omega&=&m\Omega(r_+)+\mathrm{i}\kappa_+(s-k)\, ,\qquad k=1,2,\cdots.
\end{eqnarray}
At integer differences of local Frobenius exponents, the corresponding bases can become degenerate. Thus, the locations in Eq. \eqref{eq:heun_candidate_poles} are candidate singularities; their survival as poles of the full ratio $B_{\rm out}/B_{\rm in}$ depends on the normalization and on possible cancellations. The candidate locations are independent of the angular separation constant, but this fact alone does not guarantee that every candidate singularity survives in the full scattering ratio.

For the asymptotic-amplitude boundary condition in Eq. \eqref{eq:semi_open_BC_full}, a further class of resonances approaches simple, uncanceled poles of $B_{\mathrm{in}}$ as $\mathcal R_0$ tends to zero. To derive their local behavior, we write the resonance condition \eqref{eq:qnm_condition_raw_full} as
\begin{eqnarray}\label{eq:resonanceD}
    \mathcal D(\omega)\equiv B_{\text{out}}(\omega)-\mathcal R_0\mathrm{e}^{-2\mathrm{i}\omega_{\text{h}}r_{\star\text{w}}}B_{\text{in}}(\omega)=0\, .
\end{eqnarray}
Here and below, the frequency dependence of the connection coefficients includes that of the angular eigenvalue $\lambda=\lambda_{s,\ell,m}(\omega)$. Suppose that $\omega=\omega_{\text{p}}$ is a simple, uncanceled pole of $B_{\text{in}}$, while $B_{\text{out}}$ is holomorphic and nonzero there. In its neighborhood, write
\begin{eqnarray}\label{eq:BinBoutLaurent}
    B_{\text{in}}(\omega)&=&\frac{b_{-1}}{\omega-\omega_{\text{p}}}+b_0+\mathcal O(\omega-\omega_{\text{p}})\, ,\label{eq:BinBoutLaurent_1}\\
    B_{\text{out}}(\omega)&=&c_0+c_1(\omega-\omega_{\text{p}})+\mathcal O((\omega-\omega_{\text{p}})^2)\, ,\label{eq:BinBoutLaurent_2}
\end{eqnarray}
with $b_{-1}\neq0$ and $c_0\neq0$. Define
\begin{eqnarray}
    \epsilon_{\text{p}}\equiv\mathcal R_0\mathrm{e}^{-2\mathrm{i}\omega_{\text{h},\text{p}}r_{\star\text{w}}}\, ,\qquad
    \omega_{\text{h},\text{p}}=\omega_{\text{p}}-m\Omega(r_+)\, .
\end{eqnarray}
For fixed $r_{\star\text{w}}$, expanding the phase factor gives
\begin{eqnarray}
    \mathcal D(\omega)=c_0-\frac{\epsilon_{\text{p}}b_{-1}}{\omega-\omega_{\text{p}}}+\mathcal O(\epsilon_{\text{p}}^2)\, .
\end{eqnarray}
The nearby QNM spectrum therefore satisfies
\begin{eqnarray}\label{eq:poleborn}
    \omega_{\text{pb}}(\mathcal R_0)-\omega_{\text{p}}=\epsilon_{\text{p}}\frac{b_{-1}}{c_0}+\mathcal{O}(\epsilon_{\text{p}}^2)\, .
\end{eqnarray}
Thus $\omega_{\text{pb}}\to\omega_{\text{p}}$ as $\mathcal R_0\to0^+$, even though $\omega_{\text{p}}$ is not a QNM of the perfectly open problem, for which the resonance condition is $B_{\text{out}}=0$. This nonuniform zero-reflectivity limit gives rise to the pole-born branch associated with the asymptotic-amplitude boundary prescription.

In particular, for $s=-2$, the candidate poles of $B_{\rm in}$ in Eq. \eqref{eq:heun_candidate_poles} have
\begin{eqnarray}
    \omega_{\text{p},k}=m\Omega(r_+)+\mathrm{i}\kappa_+(k-2)\, ,\qquad k=1,2,\cdots .
\end{eqnarray}
Their imaginary parts are unbounded above as $k\to\infty$. As shown in Eq. \eqref{eq:poleborn}, a pole-born branch exists near each surviving simple, uncanceled pole of $B_{\text{in}}$ satisfying the conditions stated above. If an infinite subset of these branches persists at fixed nonzero reflectivity as genuine poles of the full retarded Green function, no horizontal Laplace contour at finite height can lie above all of them. Such a spectrum would therefore be incompatible with the usual initial-value formulation based on a finite exponential growth bound.

To remove the pole-born obstruction to the initial-value formulation, we introduce a finite-radius Robin condition on the Sasaki--Nakamura (SN) variable $X$ defined in Appendix~\ref{app:sn}. Unlike the original prescription, which fixes the ratio of event-horizon-normalized Teukolsky amplitudes, this condition relates the field and its radial derivative directly at the wall. Let $r_{\text{w}}$ denote the wall radius and $r_\star(r_{\text{w}})=r_{\star\text{w}}$. We impose
\begin{eqnarray}\label{eq:robin_SN_BC}
    \mathscr{L}_{\text{w}}[X]\equiv\left(\frac{\mathrm{d}}{\mathrm{d}r_\star}+\mathrm{i}\zeta_{\text{w}}\omega_{\text{h}}\right)X\Bigg|_{r_\star=r_{\star\text{w}}}=0\, ,\qquad \zeta_{\text{w}}\geq0\, ,
\end{eqnarray}
where the real constant parameter $\zeta_{\text{w}}$ describes a local wall. A constant $\zeta_{\text{w}}$ specifies a different boundary-value problem from the fixed-$\mathcal R_0$ asymptotic-amplitude condition. Let $X_{\text{in}}^{\text{SN}}$ and $X_{\text{out}}^{\text{SN}}$ be the SN transforms (\ref{eq:app_sn_Xdef}) of $\Psi_{\text{in}}$ and $\Psi_{\text{out}}$, normalized to unit ingoing and outgoing amplitudes at the event horizon. Suppose that $X_{\text{w}}^{\text{SN}}$ is the solution of SN formulation with the Robin  condition (\ref{eq:robin_SN_BC}). Such condition fixes the corresponding basis amplitude ratio
\begin{eqnarray}\label{eq:robin_SN_rho}     
 X_{\text{w}}^{\text{SN}}&=&X_{\text{in}}^{\text{SN}}+\rho_{\text{w}}X_{\text{out}}^{\text{SN}}\, ,\qquad\text{with}\qquad\rho_{\text{w}}(\omega)=-\frac{\mathscr{L}_{\text{w}}[X_{\text{in}}^{\text{SN}}]}{\mathscr {L}_{\text{w}}[X_{\text{out}}^{\text{SN}}]
    }\, .
\end{eqnarray}
For wall placed near the event horizon, $X_{\text{in}}^{\text{SN}}\sim\mathrm{e}^{-\mathrm{i}\omega_{\text{h}}r_{\star\text{w}}}$ and $X_{\text{out}}^{\text{SN}}\sim\mathrm{e}^{+\mathrm{i}\omega_{\text{h}}r_{\star\text{w}}}$, $\rho_{\text{w}}(\omega)$ becomes
\begin{eqnarray}\label{eq:robin_SN_reflectivity}
    \rho_{\text{w}}(\omega)&\sim&\mathcal{R}_0^{\text{SN}}\mathrm{e}^{-2\mathrm{i}\omega_{\text{h}}r_{\star\text{w}}}\, ,\qquad\mathcal R_0^{\text{SN}}=\frac{1-\zeta_{\text{w}}}{1+\zeta_{\text{w}}}\, .
\end{eqnarray}
For real $\omega_{\text{h}}$, the local amplitude reflectivity has modulus at most one. A constant $\zeta_{\text{w}}$ corresponds to a constant local SN reflectivity $\mathcal R_0^{\text{SN}}$ in the near-horizon approximation. The exact finite-radius Robin condition instead determines the frequency-dependent amplitude ratio $\rho_{\text{w}}(\omega)$ through Eq. \eqref{eq:robin_SN_rho}.

The original pole-born branch persists under a fixed asymptotic reflectivity $\mathcal R_0^{\text{SN}}\sim0$ (or equivalently $\mathcal R_0\sim0$ as shown in Eq. \eqref{eq:app_sn_reflectivity_conversion}), whereas the exact finite-radius Robin condition with fixed $\zeta_{\text{w}}$ removes this pole-born branch, as shown below. We take $X_{\text{up}}^{\text{SN}}$ to be the SN transform of $\Psi_{\text{up}}$. Using the horizon amplitude transformation coefficients $\mathcal T_{\text{in}}$ and $\mathcal T_{\text{out}}$ of Appendix \ref{app:sn}, the connection relation is given by
\begin{eqnarray}\label{eq:robin_SN_connection}
    X_{\text{up}}^{\text{SN}}=\mathcal T_{\text{in}}B_{\text{in}}X_{\text{in}}^{\text{SN}}+\mathcal T_{\text{out}}B_{\text{out}}X_{\text{out}}^{\text{SN}}\, .
\end{eqnarray}
From Eq. (\ref{eq:robin_SN_rho}) and Eq. (\ref{eq:robin_SN_connection}), the exact resonance condition requires $X_{\text{up}}^{\text{SN}}$ satisfying the boundary condition
\begin{eqnarray}\label{eq:robin_SN_resonance}
    \mathcal D_{\text{w}}^{\text{SN}}(\omega)\equiv\mathscr{L}_{\text{w}}[X_{\text{up}}^{\text{SN}}]=\mathscr {L}_{\text{w}}[X_{\text{out}}^{\text{SN}}]\big(\mathcal{T}_{\text{out}}B_{\text{out}}-\rho_{\text{w}}\mathcal T_{\text{in}}B_{\text{in}}\big)=0\, .
\end{eqnarray}

Consider a simple pole $\omega_{\text{p}}$ of $B_{\text{in}}$, with expansion Eqs. \eqref{eq:BinBoutLaurent}. At the candidate poles of $B_{\text{in}}$, where $\gamma_{\text{r}}=0,-1,-2,\cdots$, the ingoing solution $\Psi_{\text{in}}$ remains regular because its Frobenius representation involves the Heun parameter $2-\gamma_{\text{r}}=2,3,4,\cdots$. The outgoing solution $\Psi_{\text{up}}$ is also regular at these frequencies. Using Eq. (\ref{eq:BinBoutLaurent_1}) and Eq. (\ref{eq:BinBoutLaurent_2}), the connection relation Eq. \eqref{eq:connection} then implies
\begin{eqnarray}
    \Psi_{\text{out}}=-\frac{b_{-1}}{c_0}\frac{\Psi_{\text{in}}(\omega_{\text{p}})}{\omega-\omega_{\text{p}}}+\mathcal{O}(1)\, .
\end{eqnarray}
Using the Teukolsky connection relation and the horizon amplitude transformation in Eq.~\eqref{eq:app_sn_horizon_waves}, with $\mathcal T_{\text{in}}$ and $\mathcal T_{\text{out}}$ given by Eqs.~\eqref{eq:app_sn_Tin} and \eqref{eq:app_sn_Tout}, respectively, we obtain from Eq.~\eqref{eq:robin_SN_connection}
\begin{eqnarray}
    X_{\text{out}}^{\text{SN}}
    =
    \frac{
    X_{\text{up}}^{\text{SN}}
    -\mathcal T_{\text{in}}B_{\text{in}}
    X_{\text{in}}^{\text{SN}}
    }{
    \mathcal T_{\text{out}}B_{\text{out}}
    }\, .
\end{eqnarray}
Substituting Eq.~\eqref{eq:BinBoutLaurent} gives\footnote{Together with Eq.~\eqref{eq:robin_SN_rho}, this implies that $\rho_\text{w}=\mathcal{O}(\omega-\omega_{\mathrm{p}})$, whereas Eq.~\eqref{eq:robin_SN_reflectivity} implies that $\rho_\text{w}$ remains nonvanishing for a fixed $\zeta_\text{w}\neq1$. This apparent contradiction arises because the latter is derived by taking the near-horizon limit prior to setting $\omega\to\omega_{\mathrm{p}}$, while the former takes $\omega\to\omega_{\mathrm{p}}$ at a fixed wall radius. These two limits do not commute.}
\begin{eqnarray}
    X_{\text{out}}^{\text{SN}}=-\frac{b_{-1}}{c_0}\frac{\mathcal{T}_{\text{in}}(\omega_{\text{p}})}{\mathcal T_{\text{out}}(\omega_{\text{p}})}\frac{X_{\text{in}}^{\text{SN}}(\omega_{\text{p}})}{\omega-\omega_{\text{p}}}+\mathcal{O}(1)\, .
\end{eqnarray}
Substituting these Laurent expansions into Eq. \eqref{eq:robin_SN_resonance} gives
\begin{eqnarray}
    \mathcal D_{\text{w}}^{\text{SN}}=\frac{\mathcal{T}_{\text{in}}(\omega_{\text{p}})b_{-1}}{\omega-\omega_{\text{p}}}\mathscr L_{\text{w}}[X_{\text{in}}^{\text{SN}}]\Big|_{\omega_{\text{p}}}-\frac{\mathcal T_{\text{in}}(\omega_{\text{p}})b_{-1}}{\omega-\omega_{\text{p}}}\mathscr{L}_{\text{w}}[X_{\text{in}}^{\text{SN}}]\Big|_{\omega_{\text{p}}}+\mathcal{O}(1)=\mathcal{O}(1)\, .
\end{eqnarray}
Thus, the poles of $\mathcal{D}_{\text{w}}^{\text{SN}}$ contributed by $B_{\text{in}}$ are canceled by those originating from $X_{\text{out}}^{\text{SN}}$ exactly. Hence $\mathcal D_{\text{w}}^{\text{SN}}$ is regular at $\omega_{\text{p}}$, and the poles of $B_{\text{in}}$ do not induce resonances. 

The original pole-born branch approaches $\omega_{\text{p}}$ as $\mathcal R_0\to0$. To examine the corresponding limit for the exact Robin condition, we take $\zeta_{\text{w}}\to1$, which corresponds to $\mathcal R_0^{\text{SN}}\to0$ in the near-horizon approximation. Unlike the near-horizon approximation, the exact Robin condition determines the frequency-dependent ratio $\rho_{\text{w}}$ through Eq. \eqref{eq:robin_SN_rho}. At an upper-half-plane frequency $\omega_{\text{p},k}$, we have $\omega_{\text{h}}=\mathrm{i}\kappa_+(k-2)$ with $k\ge3$. At the frequency $\omega_{\text{p},k}$, $X_{\text{up}}^{\text{SN}}$ is regular and its leading behavior is outgoing. Hence
\begin{eqnarray}
    X_{\text{up}}^{\text{SN}}\sim A_k\mathrm{e}^{\mathrm{i} \omega_{\text{h}} r_\star}\, ,\qquad A_k=\mathcal T_{\text{out}}(\omega_{\text{p},k}) B_{\text{out}}(\omega_{\text{p},k})\neq0\, .
\end{eqnarray}
Consequently, 
\begin{eqnarray}
    \mathcal D_{\text{w}}^{\text{SN}}(\omega_{\text{p},k},\zeta_{\text{w}}=1)\sim2\mathrm{i}\omega_{\text{h}} A_k\mathrm{e}^{\mathrm{i} \omega_{\text{h}} r_\star}\neq0
\end{eqnarray}
for a wall sufficiently close to the event horizon. By analyticity, the Robin Green function has no resonance poles in a neighborhood of $\omega_{\text{p},k}$ for sufficiently small local reflectivity $\mathcal R_0^{\text{SN}}$. The original pole-born branch is therefore absent from this neighborhood in the exact Robin problem with $\zeta_\text{w}$ fixed.

\begin{figure}[htbp]
    \centering
    \subfigure[]{\includegraphics[width=0.28\linewidth]{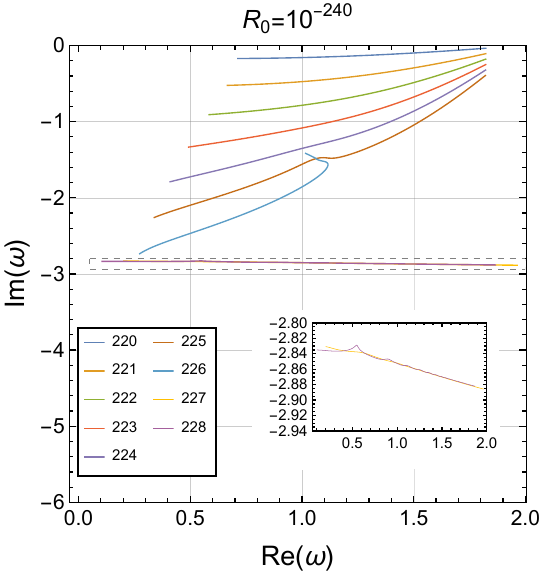}} \hfill
    \subfigure[]{\includegraphics[width=0.28\linewidth]{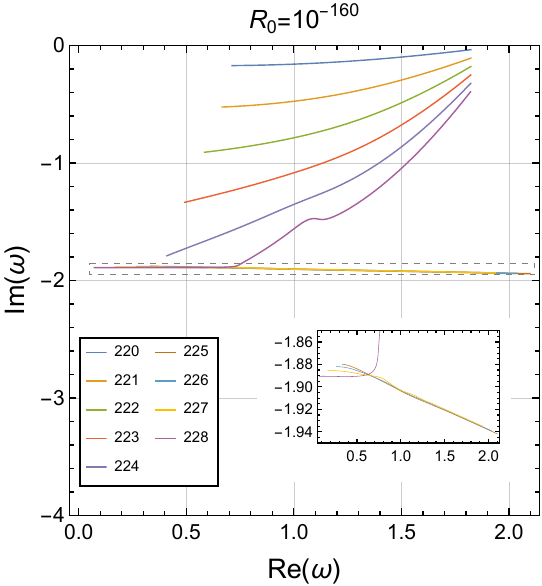}} \hfill
    \subfigure[]{\includegraphics[width=0.28\linewidth]{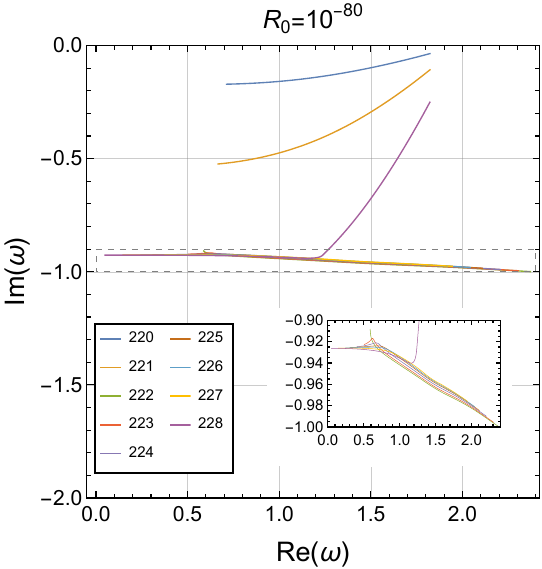}\label{fig:olofa80}}
    \caption{Rotation dependence of the selected ``+'' QNM branches at fixed wall reflectivity. In each panel, the rotation parameter varies from $a=0$ to $a=0.5$ while $\mathcal R_0$ is held fixed. Panels (a)-(c) correspond to $\mathcal{R}_0=10^{-240}$, $10^{-160}$, and $10^{-80}$, respectively. The modes are tracked by continuous variation of $a$ from the corresponding nonrotating configurations. The insets show enlarged views of the regions enclosed by the dashed boxes. }
    \label{fig:olofap}
\end{figure}

\begin{figure}[htbp]
    \centering
    \subfigure[]{\includegraphics[width=0.28\linewidth]{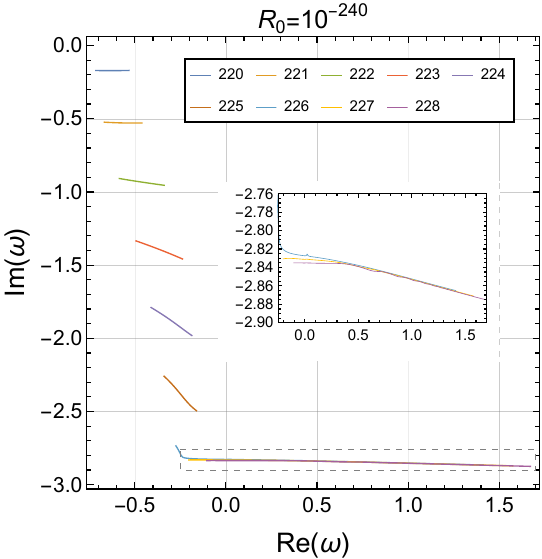}} \hfill
    \subfigure[]{\includegraphics[width=0.28\linewidth]{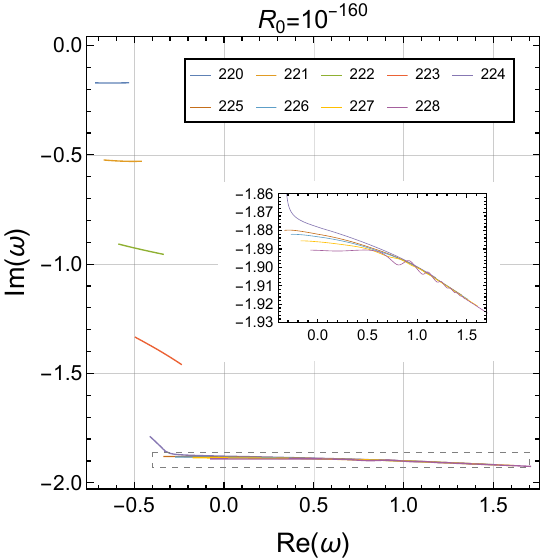}} \hfill
    \subfigure[]{\includegraphics[width=0.28\linewidth]{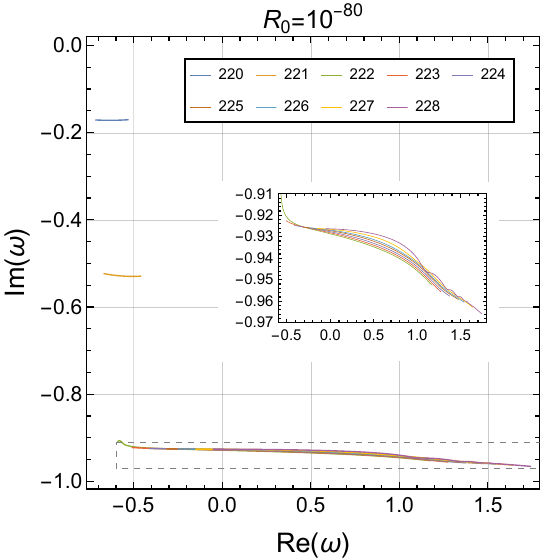}}
    \caption{Rotation dependence of the selected ``-'' QNM branches at fixed wall reflectivity. The rotation parameter varies from $a=0$ to $a=0.5$, with $\mathcal R_0$ fixed in each panel. Panels (a)-(c) correspond to $\mathcal R_0=10^{-240}$, $10^{-160}$, and $10^{-80}$, respectively. The modes are tracked by continuous variation of $a$ from the corresponding nonrotating configurations. The insets provide enlarged views of the regions enclosed by the dashed boxes.}
    \label{fig:olofam}
\end{figure}

We next vary the rotation parameter at fixed wall reflectivity. To specify the mode labels, the numerical continuation first moves from $(\mathcal{R}_0,a)=(0,0)$ to $(\mathcal{R}_0,0)$ and then increases $a$ while holding $\mathcal{R}_0$ fixed. The resulting trajectories are displayed in Figs. \ref{fig:olofap} and \ref{fig:olofam}. This path differs from that used above, and overtone assignments may change if spectral branches are exchanged. 
The trajectories in Figs. \ref{fig:olofap} and \ref{fig:olofam} show two broad responses to increasing rotation. Some ``+'' modes move toward the real axis, while other modes, including a subset of the ``-'' branch, retain a finite negative imaginary part and remain damped. This contrast is reminiscent of the distinction between damped and zero-damping families in nearly extremal Kerr spacetimes~\cite{Yang:2013qnm}, although the present modes obey different boundary conditions and are not identified with those Kerr families. 

\begin{figure}[htbp]
    \centering
    \subfigure[]{\includegraphics[width=0.35\linewidth]{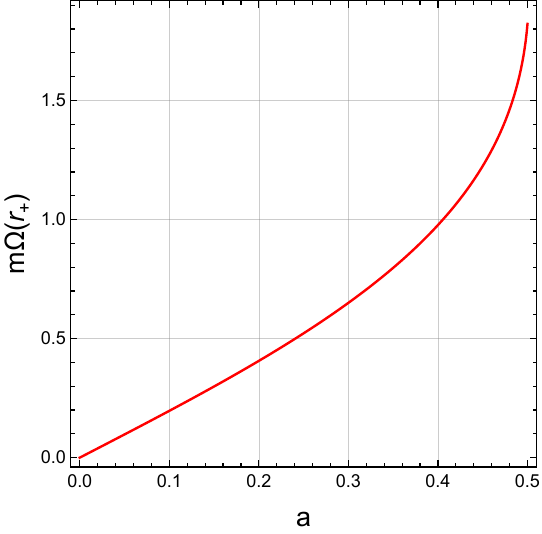}} \hspace{2cm}
    \subfigure[]{\includegraphics[width=0.35\linewidth]{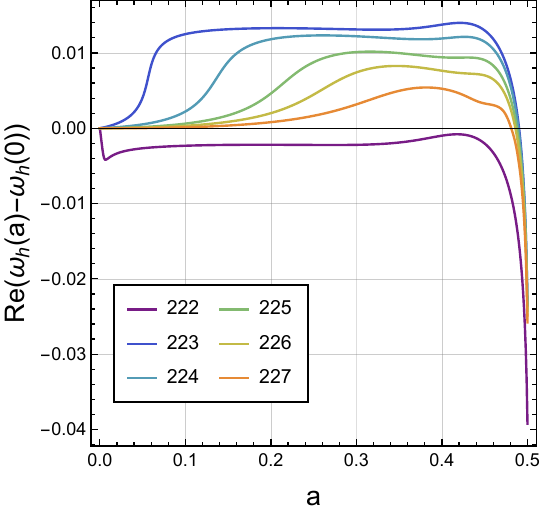}}
    \caption{The decomposition of the rotation dependence for the real parts of the selected ``+'' modes at $\mathcal R_0=10^{-80}$, corresponding to Fig. \ref{fig:olofa80}. (a): The dominate kinematic part $m\Omega(r_+)$. (b): the residual shift $\operatorname{Re}[\omega_{\text{h}}(a)-\omega_{\text{h}}(0)]=\operatorname{Re}[\omega(a)-\omega(0)-m\Omega(r_+,a)]$. The dominant part is two orders of magnitude larger than the residual part.}
    \label{fig:mOmega}
\end{figure}

In both the ``+'' and ``-'' QNM spectra, some trajectories instead display an approximately horizontal shift of their real parts with comparatively small changes in their imaginary parts. Examining Eqs. \eqref{eq:Rfunc_full} and \eqref{eq:qnm_condition_full} shows that in regions where the ratio $B_{\mathrm{out}}/B_{\mathrm{in}}$ varies relatively slowly, maintaining a nearly uniform phase of $\mathrm{e}^{2\mathrm{i}\omega_{\mathrm{h}}r_{\star\text{w}}}$ favors an approximately constant $\omega_{\mathrm{h}}$. Hence the leading horizontal displacement of $\omega$ tracks the variation of $m\Omega(r_+)$. Fig. \ref{fig:mOmega} quantifies this effect for selected ``+'' modes at $\mathcal{R}_0=10^{-80}$, corresponding to the case in Fig. \ref{fig:olofa80}, by decomposing the total frequency shift into the dominating kinematic contribution from the event-horizon angular velocity and the comparatively small residual correction $\operatorname{Re}[\omega_{\mathrm{h}}(a)-\omega_{\mathrm{h}}(0)]$. The nearly horizontal behavior of some ``-'' modes in Fig. \ref{fig:olofam} can be understood by the same phase argument, whereas other ``-'' trajectories retain finite damping rather than approaching the real axis. These shifts indicate that the wall phase is important but do not imply an exact translation law for the full spectrum.

%======================================%
%<<<<<<<<<<< Conclusions >>>>>>>>>>>>>%
%======================================%

\section{Conclusions}\label{sec:conclusions}
We have studied gravitational quasinormal modes of Kerr--de Sitter black holes in a semi-open configuration with a partially reflective inner boundary near the event horizon and an outgoing condition at the cosmological horizon. By reducing the angular and radial equations to Heun's equations, we constructed the local solutions and their connection coefficients, which allowed us to formulate the resonance condition explicitly in terms of the inner boundary reflectivity. This provides a unified framework for investigating the QNM spectrum to changes in the boundary condition and the black hole rotation.

For the gravitational sector with $s=-2$ and $\ell=m=2$, varying the reflectivity with fixed rotation produces two characteristic classes of spectral trajectories. Many modes approach the real-frequency axis as the reflectivity increases and form echo modes whose approximately regular real frequency spacing is determined primarily by the cavity length in tortoise coordinates. Other branches approach or begin with the vertical line $\operatorname{Re}\omega=m\Omega(r_+)$. As for varying rotation with fixed reflectivity, there are three different classes of spectral trajectories. In particular, for several modes, their imaginary parts remain nearly fixed, and the leading horizontal displacement of the real frequency follows the variation of $m\Omega(r_+)$, while the residual variation of the horizon frequency $\omega_{\mathrm h}$ remains comparatively small. Other modes exhibit distinct responses, with some approaching the real axis and others retaining finite damping.

The energy flux analysis establishes the relation between the Teukolsky amplitude reflectivity $\mathcal R_0$ and the corresponding ratio $\mathcal R_{\text{E}}$ of fluxes through the conserved Teukolsky-Starobinsky current. We have also derived a Sasaki--Nakamura type radial equation and constructed its conserved current, giving an alternative way to obtain the $\mathcal R_{\text{E}}$. In the Teukolsky normalization, the ratio between the flux reflectivity and the amplitude reflectivity, $\eta$,  diverges at the event horizon synchronous frequency $m\Omega(r_+)$. Consequently, even a small Teukolsky amplitude reflectivity may correspond to an active reflective wall sufficiently close to this frequency. The analysis of the crossing of the real axis reveals two distinct instability mechanisms. Outside the superradiant interval, the crossings require an active inner boundary with $\mathcal R_{\mathrm E}>1$. Inside the superradiant interval, the crossings satisfy $\mathcal R_{\mathrm E}<1$, showing that the instability is driven by the superradiance since the reflective boundary is passive here. In the nonrotating case, where no superradiant interval exists, the instability is triggered solely by the energy injection from the inner boundary.

We have further investigated the analytic structure of the resonance condition and its dependence on the boundary prescription. For the asymptotic amplitude boundary condition, a simple pole of the ingoing connection coefficient $B_{\mathrm{in}}$ generates a nearby pole-born resonance even as the reflection amplitude tends to zero. In the gravitational sector, the pole frequencies are $\omega_{\mathrm{p},k}=m\Omega(r_+)+\mathrm{i}\kappa_+(k-2)$, with positive integer $k$, and extend to arbitrarily large positive imaginary parts. If an infinite subset survives as poles of the full retarded Green function, this spectrum would obstruct the usual initial value formulation. To address this issue, we introduced an exact finite radius Robin condition directly on the Sasaki--Nakamura variable. The resulting resonance functional is regular at the poles of $B_{\mathrm{in}}$ because its singular contributions are canceled by the corresponding singular behavior of the transformed outgoing solution. In the limit of small local reflectivity, no resonances appear near these upper half-plane candidate frequencies. Thus, the pole born branches are a consequence of the asymptotic amplitude boundary prescription and are absent from the corresponding finite radius Robin problem. This result demonstrates that the spectrum depends not only on the background geometry and reflectivity, but also on how the inner boundary condition is formulated.

%======================================%
%<<<<<<<<<< Acknowledgement >>>>>>>>>>>%
%======================================%

\section*{Acknowledgement}\label{sec:acknowledgement}
This work is supported in part by the National Key R\&D Program of China Grant No.2022YFC2204603, by the National Natural Science Foundation of China with grants No.12475063, No.12075232 and No.12247103. This work is also supported by the National Natural Science Foundation of China with grants No. 12505067.

\appendix

\section{Energy-flux in Teukolsky formalism}\label{app:flux}
This appendix presents the real-frequency normalization used to compare the event-horizon Teukolsky amplitudes with flux magnitudes. The relevant relations follow from the gravitational Teukolsky--Starobinsky (TS) identities and their conserved bilinear current~\cite{Suzuki:1999tsi,Suzuki:2000abs,Nakano:2017howls}. Throughout this appendix, we consider $s=-2$ and $\omega\in\mathbb R$.

We first derive the ratio $\eta$ between the energy reflectivity $\mathcal R_{\text{E}}$ and the squared amplitude reflectivity $|\mathcal R_0|^2$ [see Eq. (\ref{eq:flux_reflectivity})]. We introduce the event-horizon quantities
\begin{eqnarray}\label{eq:app_flux_def2}
    \Sigma_{\text{h}}=r_+^2+a^2\, ,\qquad
    \mathcal K_{\text{h}}=\Xi K(r_+)=\Xi\Sigma_{\text{h}}\omega_{\text{h}}\, ,\qquad
    D_{\text{h}}=\Delta_{\text{r}}'(r_+)=2\Xi\Sigma_{\text{h}}\kappa_+ \, ,
\end{eqnarray}
where $\Xi=1+\alpha_0$ as defined in Sec. \ref{sec:background}.
The radial TS operators are given by
\begin{eqnarray}\label{eq:app_flux_Dop}
    \mathscr{D}_0=\frac{\mathrm{d}}{\mathrm{d}r}-\frac{\mathrm{i}\Xi K(r)}{\Delta_{\text{r}}}\, ,\qquad \mathscr{D}_0^\dagger=\frac{\mathrm{d}}{\mathrm{d}r}+\frac{\mathrm{i}\Xi K(r)}{\Delta_{\text{r}}}\, .
\end{eqnarray}
For a solution $R_{-2}$ of the $s=-2$ radial Teukolsky equation, the TS transformation gives a solution of the $s=+2$ equation, which we denote by
\begin{eqnarray}\label{eq:app_flux_positive_spin}
    R_{+2}\equiv\mathscr D_0^4R_{-2}\, .
\end{eqnarray}
Conversely, the reverse TS transformation operator $\Delta_{\text{r}}^2(\mathscr D_0^\dagger)^4\Delta_{\text{r}}^2$ maps a solution of the $s=+2$ equation back to a solution of the $s=-2$ equation. The composition of these two transformations satisfies the TS identity,
\begin{eqnarray}\label{eq:app_flux_TS_comp}
    \Delta_{\text{r}}^2(\mathscr D_0^\dagger)^4\Big[\Delta_{\text{r}}^2\mathscr D_0^4R_{-2}\Big]=\mathfrak S_{\text{KdS}}R_{-2}\, ,\qquad\mathfrak{S}_{\text{KdS}}\equiv|\mathcal C_{\text{KdS}}|^2\, ,
\end{eqnarray}
where $\mathcal C_{\text{KdS}}$ is the TS constant for the Kerr--de Sitter black hole. Its explicit expression is
\begin{eqnarray}\label{eq:app_CKdS_closed_form}
    |\mathcal C_{\text{KdS}}|^2&=&\widehat\lambda^2(\widehat\lambda+2)^2+8\widehat\lambda\Xi^2a\omega    \Big[(6+5\widehat\lambda)(m-a\omega)+12a\omega\Big]+144\Xi^4a^2\omega^2(m-a\omega)^2\nonumber\\
    &&-4\alpha_0\Bigl\{\widehat\lambda(\widehat\lambda+2)(\widehat\lambda-6)+12\Xi^2(m-a\omega)\Big[2m\widehat\lambda-a\omega(\widehat\lambda-6)\Big]\Bigr\}+4\alpha_0^2(\widehat\lambda-6)^2+144M^2\omega^2\Xi^2\, ,
\end{eqnarray}
where
\begin{eqnarray}
    \widehat\lambda=\lambda+4(1-\alpha_0)\, .
\end{eqnarray}

We now use the near-horizon behavior of the radial solutions to determine the TS amplitudes of $R_{+2}$. Near the event horizon,
\begin{eqnarray}
    \frac{\mathrm{d}r_{\star}}{\mathrm{d}r}\sim\frac{\Xi\Sigma_{\text{h}}}{\Delta_{\text{r}}}\, ,\qquad\frac{\mathrm{d}\Delta_{\text{r}}}{\mathrm{d}r}\sim D_{\text{h}}\, ,\qquad\Xi K(r)\sim\mathcal K_{\text{h}}\, .
\end{eqnarray}
Consequently, for $\sigma=\pm1$ and integer $p$, we have
\begin{eqnarray}
    \frac{\mathrm{d}}{\mathrm{d}r}\Big(\Delta_{\text{r}}^p\mathrm{e}^{\sigma\mathrm{i}\omega_{\text{h}}r_{\star}}\Big)\sim\big(pD_{\text{h}}+\mathrm{i}\sigma\mathcal K_{\text{h}}\big)\Delta_{\text{r}}^{p-1}\mathrm{e}^{\sigma\mathrm{i}\omega_{\text{h}}r_{\star}}\, .
\end{eqnarray}
The corresponding leading actions of the TS operators are
\begin{eqnarray}
    \mathscr{D}_0\Big(\Delta_{\text{r}}^p\mathrm{e}^{\sigma\mathrm{i}\omega_{\text{h}}r_{\star}}\Big)&\sim&\Big[pD_{\text{h}}+\mathrm{i}(\sigma-1)\mathcal K_{\text{h}}\Big]\Delta_{\text{r}}^{p-1}\mathrm{e}^{\sigma\mathrm{i}\omega_{\text{h}}r_{\star}}\label{D0}\, ,
    \\
    \mathscr{D}_0^\dagger\Big(\Delta_{\text{r}}^p\mathrm{e}^{\sigma\mathrm{i}\omega_{\text{h}}r_{\star}}\Big)&\sim&\Big[pD_{\text{h}}+\mathrm{i}(\sigma+1)\mathcal K_{\text{h}}\Big]\Delta_{\text{r}}^{p-1}\mathrm{e}^{\sigma\mathrm{i}\omega_{\text{h}}r_{\star}}\, .
\end{eqnarray}
For the ingoing wave, setting $\sigma=-1$ and applying $\mathscr D_0$ successively four times [see (\ref{D0})], with $p=2,1,0,-1$, gives
\begin{eqnarray}\label{eq:app_flux_opaction}
    \mathscr D_0^4\Big(\Delta_{\text{r}}^2\mathrm{e}^{-\mathrm{i}\omega_{\text{h}}r_{\star}}\Big)\sim q_{\text{h}}\Delta_{\text{r}}^{-2}\mathrm{e}^{-\mathrm{i}\omega_{\text{h}}r_{\star}}\, ,
\end{eqnarray}
where the above quantity $q_{\text{h}}$ is
\begin{eqnarray}\label{eq:app_flux_q}
    q_{\text{h}}=4\mathrm{i}\mathcal K_{\text{h}}(D_{\text{h}}-\mathrm{i}\mathcal K_{\text{h}})(D_{\text{h}}^2+4\mathcal K_{\text{h}}^2)\, .
\end{eqnarray}

To determine the outgoing-wave coefficient, we instead use the reverse TS transformation. Applying the adjoint operator to the outgoing horizon wave gives, 
\begin{eqnarray}\label{eq:app_flux_spins2}
    \Delta_{\text{r}}^2(\mathscr D_0^\dagger)^4\big(\Delta_{\text{r}}^2\mathrm{e}^{\mathrm{i}\omega_{\text{h}}r_{\star}}\big)\sim(2D_{\text{h}}+2\mathrm{i}\mathcal K_{\text{h}})(D_{\text{h}}+2\mathrm{i}\mathcal K_{\text{h}})(2\mathrm{i}\mathcal K_{\text{h}})(-D_{\text{h}}+2\mathrm{i}\mathcal K_{\text{h}})\mathrm{e}^{\mathrm{i}\omega_{\text{h}}r_{\star}}=q_{\text{h}}^{\star}\mathrm{e}^{\mathrm{i}\omega_{\text{h}}r_{\star}}\, .
\end{eqnarray}
Here and below, the superscript star means complex conjugate. Comparing the outgoing-wave coefficients in the TS composition identity, Eq. \eqref{eq:app_flux_TS_comp}, determines the outgoing coefficient of $R_{+2}$ to be $\mathfrak S_{\text{KdS}}/q_{\text{h}}^{\star}$. Thus, the two horizon asymptotic solutions take the following forms
\begin{eqnarray}\label{eq:app_flux_spins-2}
    R_{-2}&\sim&Z_{\text{in}}^{(\text{h})}\Delta_{\text{r}}^2\mathrm{e}^{-\mathrm{i}\omega_{\text{h}} r_{\star}}+Z_{\text{out}}^{(\text{h})}\mathrm{e}^{\mathrm{i}\omega_{\text{h}} r_{\star}}\, ,\\
    R_{+2}&\sim&q_{\text{h}}Z_{\text{in}}^{(\text{h})}\Delta_{\text{r}}^{-2}\mathrm{e}^{-\mathrm{i}\omega_{\text{h}} r_{\star}}+\frac{\mathfrak S_{\text{KdS}}}{q_{\text{h}}^{\star}}Z_{\text{out}}^{(\text{h})}\mathrm{e}^{\mathrm{i}\omega_{\text{h}} r_{\star}}\, .
\end{eqnarray}

The spin-reversed solution
$\widetilde R_{-2}=\Delta_{\text{r}}^2R_{+2}^{\star}$
also satisfies the same $s=-2$ radial equation. Since
$R_{-2}$ and $\widetilde R_{-2}$ are solutions of the radial Teukolsky equation (\ref{eq:radial_teuk}), their weighted Wronskian reads
\begin{eqnarray}\label{eq:app_flux_conserved_W}
    \mathscr{W}_{\text{TS}}\equiv\mathscr W_{\text{T}}[R_{-2},\widetilde R_{-2}]=\Delta_{\text{r}}^{-1}\Big(R_{-2}\frac{\mathrm{d}\widetilde R_{-2}}{\mathrm{d}r}-\widetilde R_{-2}\frac{\mathrm{d}R_{-2}}{\mathrm{d}r}\Big)\, ,
\end{eqnarray}
is conserved, namely $\mathrm{d}\mathscr W_{\text{TS}}/\mathrm{d}r=0$. Evaluating this Wronskian at the event horizon gives
\begin{eqnarray}\label{eq:app_flux_W_horizon}
    \mathscr W_{\text{TS},\text{h}}=2(D_{\text{h}}-\mathrm{i}\mathcal K_{\text{h}})q_{\text{h}}^{\star}\Big[\frac{\mathfrak S_{\text{KdS}}}{|q_{\text{h}}|^2}|Z_{\text{out}}^{(\text{h})}|^2-|Z_{\text{in}}^{(\text{h})}|^2\Big]\, .
\end{eqnarray}
We introduce the conserved current $\mathscr J=\mathscr W_{\text{TS}}/(2\mathrm{i})$, whose values at the event and cosmological horizons are denoted by $\mathscr J_{\text{h}}$ and $\mathscr J_{\text{c}}$, respectively. Evaluating it at the event horizon gives
\begin{eqnarray}\label{eq:app_flux_J_horizon}
    \mathscr{J}_{\text{h}}={}4\mathcal K_{\text{h}}(D_{\text{h}}^2+4\mathcal K_{\text{h}}^2)(D_{\text{h}}^2+\mathcal K_{\text{h}}^2)\Big[|Z_{\text{in}}^{(\text{h})}|^2-\frac{\mathfrak S_{\text{KdS}}}{|q_{\text{h}}|^2}|Z_{\text{out}}^{(\text{h})}|^2\Big]\, .
\end{eqnarray}
The corresponding positive traveling-wave magnitudes are defined as
\begin{eqnarray}\label{eq:app_flux_waves_current}
    \mathcal F_{\text{in}}=|\omega_{\text{h}}|\mathscr N_{\text{h}}|Z_{\text{in}}^{(\text{h})}|^2\, ,\quad \mathcal F_{\text{out}}=|\omega_{\text{h}}|\mathscr N_{\text{h}}\frac{\mathfrak S_{\text{KdS}}}{|q_{\text{h}}|^2}|Z_{\text{out}}^{(\text{h})}|^2\, ,\quad \text{with} \quad \mathscr N_{\text{h}}=4\Xi\Sigma_{\text{h}}
    (D_{\text{h}}^2+4\mathcal K_{\text{h}}^2)
    (D_{\text{h}}^2+\mathcal K_{\text{h}}^2)>0\, .
\end{eqnarray}
From Eq. (\ref{eq:app_flux_J_horizon}) and Eqs. (\ref{eq:app_flux_waves_current}), it follows that
\begin{eqnarray}
    \mathscr J_{\text{h}}=\operatorname{sgn}(\omega_{\text{h}})(\mathcal F_{\text{in}}-\mathcal F_{\text{out}})\, .
\end{eqnarray}
The identification of these positive magnitudes
with physical gravitational energy fluxes requires
matching the Teukolsky amplitudes to the radiative
metric amplitudes. For a fixed real-frequency mode,
this matching introduces a common mode-dependent
normalization factor for the ingoing and outgoing
waves, which cancels in their flux ratio $\mathcal F_{\text{out}}/\mathcal F_{\text{in}}$
~\cite{Suzuki:2000abs,Nakano:2017howls}.

We now impose the boundary condition
\begin{eqnarray}\label{eq:app_flux_wall_amp}
    \frac{Z_{\text{out}}^{(\text{h})}}{Z_{\text{in}}^{(\text{h})}}=\mathcal R_0\mathrm{e}^{-2\mathrm{i}\omega_{\text{h}}r_{\star\text{w}}}\, ,
\end{eqnarray}
under which the energy reflectivity, defined as the ratio of the outgoing to incoming positive flux magnitudes, becomes
\begin{eqnarray}\label{eq:app_flux_passive}
    \mathcal R_{\text{E}}=\frac{\mathcal F_{\text{out}}}{\mathcal F_{\text{in}}}=\eta(\omega)|\mathcal R_0|^2\, ,
\end{eqnarray}
where the ratio is
\begin{eqnarray}\label{eq:app_flux_eta}
    \eta(\omega)\equiv\frac{\mathcal R_{\text{E}}}{|\mathcal R_0|^2}=\frac{\mathfrak S_{\text{KdS}}}{|q_{\text{h}}|^2}=\frac{|\mathcal C_{\text{KdS}}|^2}{256\mathcal K_{\text{h}}^2(\mathcal K_{\text{h}}^2+D_{\text{h}}^2/4)^2(\mathcal K_{\text{h}}^2+D_{\text{h}}^2)}\, .
\end{eqnarray}
Note that $\mathcal K_{\text{h}}$ is proportional to $\omega_{\text{h}}$, thus $\eta(\omega)$ diverge at $\omega_{\text{h}}=0$. For real frequencies approaching $m\Omega(r_+)$, one has $\mathcal K_{\text{h}}=\Xi\Sigma_{\text{h}}\omega_{\text{h}}\to0$. From Eq. \eqref{eq:app_flux_q}, we obtain
\begin{eqnarray}
    |q_{\text{h}}|^2\sim16(\Xi\Sigma_{\text{h}})^2D_{\text{h}}^6\omega_{\text{h}}^2\, .
\end{eqnarray}
Using $\mathcal R_{\text{E}}=\eta(\omega)|\mathcal R_0|^2$ and $\eta(\omega)=\mathfrak S_{\text{KdS}}/|q_{\text{h}}|^2$, we obtain
\begin{eqnarray}\label{eq:app_sn_sync}
    \mathcal{R}_{\text{E}}(\omega)\sim\frac{\mathfrak S_{\text{KdS}}\bigl(m\Omega(r_+)\bigr)}{16D_{\text{h}}^6(\Xi\Sigma_{\text{h}})^2[\omega-m\Omega(r_+)]^2}|\mathcal R_0|^2\, .
\end{eqnarray}
A fixed nonzero Teukolsky-amplitude reflectivity may therefore implies a active reflective wall, i.e., $\mathcal{R}_{\text{E}}>1$ as the frequency approaching $\omega=m\Omega(r_+)$.

In the $\Lambda\to0$ limit (Kerr case), our separation constant is related to the Kerr variable $Q$ used in Ref. \cite{Nakano:2017howls} by $Q=\lambda+6$. In terms of $Q$, the Kerr TS constant is
\begin{eqnarray}\label{eq:app_flux_CKerr_explicit}
    |\mathcal C_{\text{Kerr}}|^2&=&\Big(Q^2+4a\omega m-4a^2\omega^2\Big)\times\Big[(Q-2)^2+36a\omega m-36a^2\omega^2\Big]\nonumber\\
    &&+48(2Q-1)(2a^2\omega^2-a\omega m)+144\omega^2(M^2-a^2)\, .
\end{eqnarray}
The horizon quantities reduce to
\begin{eqnarray}\label{eq:app_flux_Kerr_horizon}
    \mathcal{K}_{\text{h}}\longrightarrow\Sigma_{\text{h}}\omega_{\text{h}}\, ,\qquad D_{\text{h}}\longrightarrow r_+-r_{-}=2\Sigma_{\text{h}}\kappa_{+}\, ,
\end{eqnarray}
and hence the ratio in Kerr black hole is
\begin{eqnarray}\label{eq:app_flux_etaKerr}
    \eta_{\text{Kerr}}=\frac{|\mathcal C_{\text{Kerr}}|^2}{256\Sigma_{\text{h}}^8\omega_{\text{h}}^2(\omega_{\text{h}}^2+\kappa_+^2)^2(\omega_{\text{h}}^2+4\kappa_+^2)}\, .
\end{eqnarray}
This recovers the standard Kerr results~\cite{Nakano:2017howls,Chen:2021tfc}.

Finally, we use the same conserved TS current to establish the flux relation between the event and cosmological horizons [see Eq. (\ref{eq:barrier_current})]. At the cosmological horizon, define
\begin{eqnarray}\label{eq:app_flux_cosmological_horizon}
    \Sigma_{\text{c}}=(r'_+)^2+a^2\, ,\qquad \mathcal{K}_{\text{c}}=\Xi\Sigma_{\text{c}}\omega_{\text{c}}\, ,\qquad D_{\text{c}}&=\Delta_{\text{r}}'(r'_+)\, .
\end{eqnarray}
The corresponding horizon factors are
\begin{eqnarray}\label{eq:app_flux_cosmological_weights}
    q_{\text{c}}=4\mathrm{i}\mathcal K_{\text{c}}(D_{\text{c}}-\mathrm{i}\mathcal K_{\text{c}})(D_{\text{c}}^2+4\mathcal K_{\text{c}}^2)\, ,\quad\eta_{\text{c}}=\frac{\mathfrak{S}_{\text{KdS}}}{|q_{\text{c}}|^2}\, ,\quad \text{with}\quad \mathscr N_{\text{c}}=4\Xi\Sigma_{\text{c}}(D_{\text{c}}^2+4\mathcal K_{\text{c}}^2)(D_{\text{c}}^2+\mathcal K_{\text{c}}^2)\, .
\end{eqnarray}
The corresponding event-horizon quantities are $\eta$ and $\mathscr N_{\text{h}}$, as given in Eqs. \eqref{eq:app_flux_eta} and \eqref{eq:app_flux_waves_current}. Let $Z_{\text{in}}^{(\text{c})}$ and $Z_{\text{out}}^{(\text{c})}$ denote the local Teukolsky coefficients at the cosmological horizon. The outgoing cosmological-horizon condition sets $Z_{\text{in}}^{(\text{c})}=0$. We then introduce the current-normalized amplitudes
\begin{eqnarray}\label{eq:app_flux_normalized_two_horizon}
    A_{\rightarrow}=(\mathscr{N}_{\text{h}}\eta)^{1/2}Z_{\text{out}}^{(\text{h})}\, ,\quad A_{\leftarrow}=\mathscr N_{\text{h}}^{1/2}Z_{\text{in}}^{(\text{h})}\, ,\quad T=(\mathscr N_{\text{c}}\eta_{\text{c}})^{1/2}Z_{\text{out}}^{(\text{c})}\, .
\end{eqnarray}
Evaluating the same conserved current at the two horizons gives
\begin{eqnarray}\label{eq:app_flux_current_scattering}
    \mathscr{J}_{\text{h}}&=&\omega_{\text{h}}\mathscr N_{\text{h}}\Big(|Z_{\text{in}}^{(\text{h})}|^2-\eta|Z_{\text{out}}^{(\text{h})}|^2\Big)=\omega_{\text{h}}\Big(|A_{\leftarrow}|^2-|A_{\rightarrow}|^2\Big)\, ,\\
    \mathscr{J}_{\text{c}}&=&\omega_{\text{c}}\mathscr N_{\text{c}}\Big(|Z_{\text{in}}^{(\text{c})}|^2-\eta_{\text{c}}|Z_{\text{out}}^{(\text{c})}|^2\Big)=-\omega_{\text{c}}|T|^2\, .
\end{eqnarray}
Since the current is conserved, $\mathscr J_{\text{h}}=\mathscr J_{\text{c}}$, and we obtain Eq. \eqref{eq:barrier_current}, which is the current relation used in Sec. \ref{sec:qnm_spectra}. The derivation is based on the conserved TS current, whose interpretation as physical gravitational energy flux retains the qualification stated above.

\section{A Sasaki--Nakamura transformation and horizon amplitude normalization}\label{app:sn}
In this appendix, we derive the Sasaki--Nakamura (SN) radial equation. We then relate the Wronskian in the transformed variables to the conserved TS current and flux normalization established in Appendix \ref{app:flux}.

For convenience, write $\mathcal K(r)=\Xi K(r)$, where $K(r)$ is defined in Sec. \ref{sec:background}. From Eq. (\ref{eq:radial_teuk}), the $s=-2$ radial equation can then be written as
\begin{eqnarray}\label{eq:app_sn_teuk}
    \Delta_{\text{r}}R_{-2}''-\Delta_{\text{r}}'R_{-2}'-VR_{-2}=0\, ,\qquad V=-\frac{\mathcal K^2+2\mathrm{i}\mathcal K\Delta_{\text{r}}'}{\Delta_{\text{r}}}+4\mathrm{i}\mathcal K^{\prime}+\lambda+4(1-\alpha_0)+2\Lambda r^2\, .
\end{eqnarray}
Here and below, primes denote differentiation with respect to $r$. The transformed radial function is defined by
\begin{eqnarray}\label{eq:app_sn_Xdef}
    X=\frac{\sqrt{r^2+a^2}}{\Delta_{\text{r}}}\Big(AR_{-2}+B\mathscr D_0R_{-2}\Big)\, ,
\end{eqnarray}
with two functions $A$ and $B$ given by
\begin{eqnarray}\label{eq:app_sn_AB}
    A=3\mathrm{i}\mathcal K'+\lambda+4(1-\alpha_0)+2\Lambda r^2+\frac{6\Delta_{\text{r}}}{r^2}\, ,\quad B=-2\mathrm{i}\mathcal K+\Delta_{\text{r}}'-\frac{4\Delta_{\text{r}}}{r}\, ,
\end{eqnarray}
where $\mathscr D_0$ is given in Eq. \eqref{eq:app_flux_Dop}. According to Eq. \eqref{eq:app_sn_teuk} and  Eq. \eqref{eq:app_sn_Xdef}, we have
\begin{eqnarray}\label{eq:app_sn_Xp}
    X^{\prime}=\frac{\sqrt{r^2+a^2}}{\Delta_r}\Big(C_{\rm SN}R_{-2}+D_{\rm SN}R_{-2}'\Big)\, ,
\end{eqnarray}
where
\begin{eqnarray}\label{eq:app_sn_CD}
    C_{\text{SN}}(r)=\Big(A-\frac{\mathrm{i}\mathcal K B}{\Delta_{\text{r}}}\Big)'+\frac{VB}{\Delta_{\text{r}}}+\Big(\frac{r}{r^2+a^2}-\frac{\Delta_{\text{r}}'}{\Delta_{\text{r}}}\Big)\Big(A-\frac{\mathrm{i}\mathcal K B}{\Delta_{\text{r}}}\Big)\, ,\quad D_{\text{SN}}(r)=A-\frac{\mathrm{i}\mathcal K B}{\Delta_{\text{r}}}+B'+\frac{rB}{r^2+a^2}\, .
\end{eqnarray}
Eqs. \eqref{eq:app_sn_Xdef} and \eqref{eq:app_sn_Xp} is a linear map $\mathcal T_{\text{SN}}$ from $(R_{-2}, R^{\prime}_{-2})$ to $(X, X')$, whose determinant is 
\begin{eqnarray}
   \frac{r^2+a^2}{\Delta^2_{\text{r}}} \gamma(r)\, ,
\end{eqnarray}
with
\begin{eqnarray}\label{eq:app_sn_gamma}
    \gamma(r)=\Big(A-\frac{\mathrm{i}\mathcal K B}{\Delta_{\text{r}}}\Big)D_{\text{SN}}-BC_{\text{SN}}=A\Big(A+B^{\prime}+\frac{4B}{r}\Big)-BA^{\prime}+\frac{6B^2}{r^2}\, .
\end{eqnarray}
The resulting SN radial equation takes the form
\begin{eqnarray} \label{eq:app_sn_master}
    \frac{\mathrm{d}^2X}{\mathrm{d}r_{\star}^2}-F_{\text{SN}}(r)\frac{\mathrm{d}X}{\mathrm{d}r_{\star}}-U_{\text{SN}}(r)X=0\, ,
\end{eqnarray}
where
\begin{eqnarray}\label{eq:app_sn_F}
    F_{\text{SN}}(r)=\frac{\Delta_{\text{r}}}{\Xi(r^2+a^2)}\frac{\gamma'(r)}{\gamma(r)}\, ,
\end{eqnarray}
and
\begin{eqnarray}\label{eq:app_sn_U}
    U_{\text{SN}}(r)=\frac{\Delta_{\text{r}}^2}{\Xi^2(r^2+a^2)^2\gamma(r)}\Bigg\{D_{\text{SN}}\Big[C_{\text{SN}}'+\Big(\frac{r}{r^2+a^2}-\frac{\Delta_{\text{r}}'}{\Delta_{\text{r}}}\Big)C_{\text{SN}}+\frac{V}{\Delta_{\text{r}}}D_{\text{SN}}\Big]-C_{\text{SN}}\Big[C_{\text{SN}}+D_{\text{SN}}'+\frac{r}{r^2+a^2}D_{\text{SN}}\Big]\Bigg\}\, .
\end{eqnarray}
Equations \eqref{eq:app_sn_teuk}-\eqref{eq:app_sn_U} specify the SN radial equation corresponding to the chosen transformation. At the event horizon, we use the quantities $\Sigma_{\text{h}}$, $\mathcal K_{\text{h}}$ and $D_{\text{h}}$ defined in Appendix \ref{app:flux}, and introduce
\begin{eqnarray}\label{eq:app_sn_horizon_def}
    A_+=\lambda+4(1-\alpha_0)+2\Lambda r_+^2+3\mathrm{i}\mathcal K'(r_+)\, .
\end{eqnarray}
An explicit expression for $\gamma(r_+)$ is
\begin{eqnarray}\label{eq:app_sn_gamma_h}
    \gamma(r_+)=A_+\Big[A_{+}+\Delta_{\text{r}}''(r_+)-2\mathrm{i}\mathcal K'(r_+)-\frac{8\mathrm{i}\mathcal K_{\text{h}}}{r_+}\Big]-(D_{\text{h}}-2\mathrm{i}\mathcal K_{\text{h}})\Big[3\mathrm{i}\mathcal K''(r_+)+4\Lambda r_++\frac{12\mathrm{i}\mathcal K_{\text{h}}}{r_+^2}\Big]\, .
\end{eqnarray}

For a solution with the Teukolsky asymptotics in Eq. \eqref{eq:app_flux_spins-2}, the transformed solution behaves as
\begin{eqnarray}\label{eq:app_sn_horizon_waves}
    X\sim\mathcal T_{\text{in}}Z_{\text{in}}^{\text{(h)}}\mathrm{e}^{-\mathrm{i}\omega_{\text{h}}r_{\star}}+\mathcal T_{\text{out}}Z_{\text{out}}^{\text{(h)}}\mathrm{e}^{\mathrm{i}\omega_{\text{h}}r_{\star}}\, ,\quad r\to r_{+}\, ,
\end{eqnarray}
where $Z_{\text{in}}^{\text{(h)}}$ and $Z_{\text{out}}^{\text{(h)}}$ are precisely those defined in Eq. \eqref{eq:app_flux_spins-2}. Note that $X$ behaves as plane waves near the boundaries, which is in fact the starting point of the SN formalism. For the ingoing wave, substituting the near-horizon asymptotics in Eq. \eqref{eq:app_flux_spins-2} into the SN transformation Eq. \eqref{eq:app_sn_Xdef}, we obtain
\begin{eqnarray}\label{eq:app_sn_Tin}
    \mathcal T_{\text{in}}=2\sqrt{\Sigma_{\text{h}}}(D_{\text{h}}-\mathrm{i}\mathcal K_{\text{h}})(D_{\text{h}}-2\mathrm{i}\mathcal K_{\text{h}})\, .
\end{eqnarray}
To determine the outgoing coefficient $\mathcal{T}_{\text{out}}$, we define the Wronskian for two solutions $X^{(1)}$ and $X^{(2)}$ of Eq. \eqref{eq:app_sn_master} as
\begin{eqnarray}
    \mathscr W_{\text{SN}}[X^{(1)},X^{(2)}]\equiv\frac{\Xi}{\gamma(r)}\Big(X^{(1)}\frac{\mathrm{d}X^{(2)}}{\mathrm{d}r_{\star}}-\frac{\mathrm{d}X^{(1)}}{\mathrm{d}r_{\star}}X^{(2)}\Big)\, .
\end{eqnarray}
We have the following identity 
\begin{eqnarray}\label{eq:app_sn_W_identity}
    \mathscr W_{\text{SN}}[X^{(1)},X^{(2)}]=\frac{1}{\Delta_{\text{r}}}\Big(R^{(1)}\frac{\mathrm{d}R^{(2)}}{\mathrm{d}r}-\frac{\mathrm{d}R^{(1)}}{\mathrm{d}r}R^{(2)}\Big)=\mathscr W_{\text{T}}[R^{(1)},R^{(2)}]\, ,
\end{eqnarray}
where $R^{(1)}$ and $R^{(2)}$, defined as $(R^{(1)},R^{(1)\prime})\equiv\mathcal{T}_{\text{SN}}^{-1}(X^{(1)},X^{(1)\prime})$ and $(R^{(2)},R^{(2)\prime})\equiv\mathcal T_{\text{SN}}^{-1}(X^{(2)},X^{(2)\prime})$, are two independent solutions of the radial equation, whose near-horizon asymptotics are
\begin{eqnarray}
    R^{(1)}\sim Z_{\text{in}}^{(\text{h})}\Delta_{\text{r}}^2\mathrm{e}^{-\mathrm{i}\omega_{\text{h}} r_{\star}}\, ,\qquad R^{(2)}\sim Z_{\text{out}}^{(\text{h})}\mathrm{e}^{\mathrm{i}\omega_{\text{h}} r_{\star}}\, .
\end{eqnarray}
Together with the corresponding SN asymptotics $X^{(1)}\sim\mathcal T_{\text{in}}Z_{\text{in}}^{\text{(h)}}\mathrm{e}^{-\mathrm{i}\omega_{\rm h}r_{\star}}$ and $X^{(2)}\sim\mathcal T_{\text{out}}Z_{\text{out}}^{\text{(h)}}\mathrm{e}^{\mathrm{i}\omega_{\rm h}r_{\star}}$, we obtain
\begin{eqnarray}\label{eq:app_sn_Tout}
    \mathcal T_{\text{out}}=-\frac{\gamma(r_+)\sqrt{\Sigma_{\text{h}}}}{2\mathrm{i}\mathcal K_{\text{h}}(D_{\text{h}}-2\mathrm{i}\mathcal K_{\text{h}})}\, .
\end{eqnarray}
Consequently, the Teukolsky and SN amplitude reflectivities are related by
\begin{eqnarray}\label{eq:app_sn_reflectivity_conversion}
    \mathcal R_0^{\text{SN}}\equiv\frac{\mathcal T_{\text{out}}}{\mathcal T_{\text{in}}}\mathcal R_0=\frac{\gamma(r_+)}{4\mathcal K_{\text{h}}(2\mathcal K_{\text{h}}+\mathrm{i}D_{\text{h}})^2(\mathcal K_{\text{h}}+\mathrm{i}D_{\text{h}})}\mathcal R_0\, .
\end{eqnarray}
Here $\mathcal{R}_0$ is the Teukolsky amplitude reflectivity used in the main text, whereas $\mathcal{R}_0^{\text{SN}}$ denotes the corresponding amplitude reflectivity in the SN normalization. We next construct the conserved current directly from the SN equation.

Similar to Appendix \ref{app:flux}, on any radial interval where $\gamma(r)\neq0$, we can introduce the conserved SN bilinear current
\begin{eqnarray}\label{eq:app_sn_current}
    \mathscr J_{\text{SN}}=\frac{1}{2\mathrm{i}}\mathscr W_{\text{SN}}[X^{(1)},X^{(2)}]\, ,\qquad \frac{\mathrm{d}\mathscr{J}_{\text{SN}}}{\mathrm{d}r_{\star}}=0\, .
\end{eqnarray}
For real frequencies, the coefficients of the SN equation are generally complex, and $X^{\star}$ is not necessarily another solution of the same equation. We therefore construct a second solution using the spin-reversed Teukolsky solution. In particular, the TS transformation and its composition identity give
\begin{eqnarray}\label{eq:app_sn_paired_solution}
    R_{+2}=\mathscr{D}_0^4R_{-2}\, ,\quad \widetilde R_{-2}=\Delta_{\text{r}}^2R_{+2}^{\star}\, ,\qquad\widetilde X=\mathcal T_{\text{SN}}(\widetilde R_{-2},\widetilde R_{-2}^{\prime})\, ,
\end{eqnarray}
where $\mathcal T_{\text{SN}}(R,R^{\prime})$ denotes the transformation defined in Eq. \eqref{eq:app_sn_Xdef}. For real $\omega$, $\widetilde R_{-2}$ is also a solution of the $s=-2$ radial equation, and hence $\widetilde X$ solves the same SN equation as $X$. Using the horizon transformation coefficients $\mathcal T_{\text{in}}$ and $\mathcal T_{\text{out}}$ obtained above together with Eq. \eqref{eq:app_flux_spins-2}, we have
\begin{eqnarray}\label{eq:app_sn_paired_waves}
    \widetilde X\sim\mathcal T_{\text{in}}\frac{\mathfrak S_{\text{KdS}}}{q_{\text{h}}}Z_{\text{out}}^{(\text{h})\star}\mathrm{e}^{-\mathrm{i}\omega_{\text{h}}r_{\star}}+\mathcal T_{\text{out}}q_{\text{h}}^{\star}Z_{\text{in}}^{(\text{h})\star}\mathrm{e}^{\mathrm{i}\omega_{\text{h}}r_{\star}}\, .
\end{eqnarray}
Substituting Eqs. \eqref{eq:app_sn_horizon_waves} and \eqref{eq:app_sn_paired_waves} into the SN current, Eq. \eqref{eq:app_sn_current}, and evaluating it at the event horizon, we find
\begin{eqnarray}\label{eq:app_sn_current_horizon_1}
    \mathscr J_{\text{SN,h}}&=&\frac{\Xi\omega_{\text{h}}\mathcal T_{\text{in}}\mathcal T_{\text{out}}}{\gamma(r_+)}\Big[q_{\text{h}}^{\star}|Z_{\text{in}}^{(\text{h})}|^2-\frac{\mathfrak S_{\text{KdS}}}{q_{\text{h}}}|Z_{\text{out}}^{(\text{h})}|^2\Big]\nonumber\\
    &=&4\mathcal K_{\text{h}}(D_{\text{h}}^2+\mathcal K_{\text{h}}^2)(D_{\text{h}}^2+4\mathcal K_{\text{h}}^2)|Z_{\text{in}}^{(\text{h})}|^2-\frac{\mathfrak S_{\text{KdS}}}{4\mathcal K_{\text{h}}(D_{\text{h}}^2+4\mathcal K_{\text{h}}^2)}|Z_{\text{out}}^{(\text{h})}|^2\, .
\end{eqnarray}
We now express this result (\ref{eq:app_sn_current_horizon_1}) entirely in terms of the SN amplitudes. From Eqs. \eqref{eq:app_sn_Tin} and \eqref{eq:app_sn_Tout}, we have
\begin{eqnarray}\label{eq:app_sn_Tnorms}
    |\mathcal T_{\text{in}}|^2=4\Sigma_{\text{h}}(D_{\text{h}}^2+\mathcal K_{\text{h}}^2)(D_{\text{h}}^2+4\mathcal K_{\text{h}}^2)\, ,\quad |\mathcal{T}_{\text{out}}|^2=\frac{\Sigma_{\text{h}}|\gamma(r_+)|^2}{4\mathcal K_{\text{h}}^2(D_{\text{h}}^2+4\mathcal K_{\text{h}}^2)}\, .
\end{eqnarray}
Introducing the SN horizon amplitudes
\begin{eqnarray}\label{eq:amplitude}
    \mathcal T_{\text{in}}B_{\text{in}}=\mathcal T_{\text{in}}Z_{\text{in}}^{(\text{h})}\, ,\qquad\mathcal T_{\text{out}}B_{\text{out}}=\mathcal T_{\text{out}}Z_{\text{out}}^{(\text{h})}\, ,
\end{eqnarray}
Eq. \eqref{eq:app_sn_current_horizon_1} can be rewritten as
\begin{eqnarray}\label{eq:app_sn_current_SN}
    \mathscr J_{\text{SN,h}}=\operatorname{sgn}(\omega_{\text{h}})(\mathcal F_{\text{in}}^{\text{SN}}-\mathcal F_{\text{out}}^{\text{SN}})\, ,
\end{eqnarray}
where the positive traveling-wave magnitudes in the SN normalization are
\begin{eqnarray}\label{eq:app_sn_fluxes}
    \mathcal F_{\text{in}}^{\text{SN}}=|\omega_{\text{h}}|\Xi|\mathcal T_{\text{in}}B_{\text{in}}|^2\, ,\qquad \mathcal F_{\text{out}}^{\text{SN}}=|\omega_{\text{h}}|\Xi\frac{\mathfrak S_{\text{KdS}}}{|\gamma(r_+)|^2}|\mathcal T_{\text{out}}B_{\text{out}}|^2\, .
\end{eqnarray}
For a fixed real-frequency mode, the conversion from this conserved current to the radiative energy flux supplies a common mode-dependent normalization for the two traveling waves, which cancels in their ratio. Therefore, the energy reflectivity in the SN normalization is
\begin{eqnarray}\label{eq:app_sn_eta_SN}
    \mathcal R_{\text{E}}\equiv\frac{\mathcal F_{\text{out}}^{\text{SN}}}{\mathcal F_{\text{in}}^{\text{SN}}}=\eta_{\text{SN}}(\omega)|\mathcal R_0^{\text{SN}}|^2\, ,\qquad\text{with}\qquad\eta_{\text{SN}}(\omega)=\frac{\mathfrak S_{\text{KdS}}}{|\gamma(r_+)|^2}=\frac{|\mathcal C_{\text{KdS}}|^2}{|\gamma(r_+)|^2}\, .
\end{eqnarray}
Here we have used
\begin{eqnarray}
    \frac{\mathcal T_{\text{out}}B_{\text{out}}}{\mathcal T_{\text{in}}B_{\text{in}}}=\mathcal R_0^{\text{SN}}\mathrm{e}^{-2\mathrm{i}\omega_{\text{h}}r_{\star\text{w}}}\, ,\qquad\Bigg|\frac{\mathcal T_{\text{out}}B_{\text{out}}}{\mathcal T_{\text{in}}B_{\text{in}}}\Bigg|=|\mathcal R_0^{\text{SN}}|\, ,
\end{eqnarray}
which follows from the boundary condition in the SN formalism. Thus, with the relation between the Teukolsky and SN amplitudes Eqs. \eqref{eq:amplitude}, we can derive the TS $\eta$ independently using SN formalism. 

A useful consistency check is provided by the Schwarzschild--de Sitter (SdS) limit. At $a=0$, the transformed equation reduces to the Regge--Wheeler equation
\begin{eqnarray}\label{eq:app_sn_SdS_RW}
    \frac{\mathrm{d}^2X}{\mathrm{d}r_{\star}^2}+\Big[\omega^2-V_{\text{RW}}(r)\Big]X=0\, ,\qquad V_{\text{RW}}(r)=f(r)\Big[\frac{\ell(\ell+1)}{r^2}-\frac{6M}{r^3}\Big]\, ,\qquad f(r)=1-\frac{2M}{r}-\frac{\Lambda r^2}{3}\, .
\end{eqnarray}
In this limit, $\gamma(r)$ is independent of $r$, so that $F_{\text{SN}}=0$ and $U_{\text{SN}}=V_{\text{RW}}-\omega^2$. In particular,
\begin{eqnarray}\label{eq:app_sn_SdS_gamma}
    \gamma(r_+)=(\lambda+4)(\lambda+6)-12\mathrm{i}M\omega\, ,\qquad|\gamma(r_+)|^2=\left.\mathfrak S_{\text{KdS}}\right|_{a=0}\, .
\end{eqnarray}
With $\lambda=\ell(\ell+1)-6$, the last identity follows from the SdS limit of the TS composition constant. Combining Eqs. \eqref{eq:app_sn_eta_SN} and \eqref{eq:app_sn_SdS_gamma}, we obtain
\begin{eqnarray}
    \eta_{\text{SN}}^{\text{SdS}}=1\, .
\end{eqnarray}
This agrees with the standard unit-amplitude flux normalization of the Regge--Wheeler equation with a real potential. Thus, in the SdS limit, $\mathcal R_{\text{E}}=|\mathcal R_0^{\text{SN}}|^2$.

\bibliography{mainreference}

@article{Ashida:2020dkc,
    author = "Ashida, Yuto and Gong, Zongping and Ueda, Masahito",
    title = "{Non-Hermitian physics}",
    eprint = "2006.01837",
    archivePrefix = "arXiv",
    primaryClass = "cond-mat.mes-hall",
    doi = "10.1080/00018732.2021.1876991",
    journal = "Adv. Phys.",
    volume = "69",
    number = "3",
    pages = "249--435",
    year = "2021"
}

@article{Berti:2009kk,
    author = "Berti, Emanuele and Cardoso, Vitor and Starinets, Andrei O.",
    title = "{Quasinormal modes of black holes and black branes}",
    eprint = "0905.2975",
    archivePrefix = "arXiv",
    primaryClass = "gr-qc",
    doi = "10.1088/0264-9381/26/16/163001",
    journal = "Class. Quant. Grav.",
    volume = "26",
    pages = "163001",
    year = "2009"
}

@article{Berti:2022xfj,
    author = "Berti, Emanuele and Cardoso, Vitor and Cheung, Mark Ho-Yeuk and Di Filippo, Francesco and Duque, Francisco and Martens, Paul and Mukohyama, Shinji",
    title = "{Stability of the fundamental quasinormal mode in time-domain observations against small perturbations}",
    eprint = "2205.08547",
    archivePrefix = "arXiv",
    primaryClass = "gr-qc",
    doi = "10.1103/PhysRevD.106.084011",
    journal = "Phys. Rev. D",
    volume = "106",
    number = "8",
    pages = "084011",
    year = "2022"
}

@article{Berti:2025hly,
    author = "Berti, Emanuele and others",
    title = "{Black hole spectroscopy: from theory to experiment}",
    eprint = "2505.23895",
    archivePrefix = "arXiv",
    primaryClass = "gr-qc",
    month = "5",
    year = "2025"
}

@article{Boyanov:2023qqf,
    author = "Boyanov, Valentin and Cardoso, Vitor and Destounis, Kyriakos and Jaramillo, Jos{\'e} Luis and Panosso Macedo, Rodrigo",
    title = "{Structural aspects of the anti{\textendash}de Sitter black hole pseudospectrum}",
    eprint = "2312.11998",
    archivePrefix = "arXiv",
    primaryClass = "gr-qc",
    doi = "10.1103/PhysRevD.109.064068",
    journal = "Phys. Rev. D",
    volume = "109",
    number = "6",
    pages = "064068",
    year = "2024"
}

@article{Boyanov:2024fgc,
    author = "Boyanov, Valentin",
    title = "{On destabilising quasi-normal modes with a radially concentrated perturbation}",
    eprint = "2410.11547",
    archivePrefix = "arXiv",
    primaryClass = "gr-qc",
    doi = "10.3389/fphy.2024.1511757",
    journal = "Front. in Phys.",
    volume = "12",
    pages = "1511757",
    year = "2024"
}

@article{Bueno:2017hyj,
    author = "Bueno, Pablo and Cano, Pablo A. and Goelen, Frederik and Hertog, Thomas and Vercnocke, Bert",
    title = "{Echoes of Kerr-like wormholes}",
    eprint = "1711.00391",
    archivePrefix = "arXiv",
    primaryClass = "gr-qc",
    doi = "10.1103/PhysRevD.97.024040",
    journal = "Phys. Rev. D",
    volume = "97",
    number = "2",
    pages = "024040",
    year = "2018"
}

@article{Cai:2025irl,
    author = "Cai, Rong-Gen and Cao, Li-Ming and Chen, Jia-Ning and Guo, Zong-Kuan and Wu, Liang-Bi and Zhou, Yu-Sen",
    title = "{Pseudospectrum for the Kerr black hole with spin s=0 case}",
    eprint = "2501.02522",
    archivePrefix = "arXiv",
    primaryClass = "gr-qc",
    reportNumber = "ICTS-USTC/PCFT-25-02",
    doi = "10.1103/PhysRevD.111.084011",
    journal = "Phys. Rev. D",
    volume = "111",
    number = "8",
    pages = "084011",
    year = "2025"
}

@article{Cao:2024oud,
    author = "Cao, Li-Ming and Chen, Jia-Ning and Wu, Liang-Bi and Xie, Libo and Zhou, Yu-Sen",
    title = "{The pseudospectrum and spectrum (in)stability of quantum corrected Schwarzschild black hole}",
    eprint = "2401.09907",
    archivePrefix = "arXiv",
    primaryClass = "gr-qc",
    reportNumber = "ICTS-USTC/PCFT-24-03",
    doi = "10.1007/s11433-024-2435-5",
    journal = "Sci. China Phys. Mech. Astron.",
    volume = "67",
    number = "10",
    pages = "100412",
    year = "2024"
}

@article{Cao:2024sot,
    author = "Cao, Li-Ming and Wu, Liang-Bi and Zhou, Yu-Sen",
    title = "{The (in)stability of quasinormal modes of Boulware-Deser-Wheeler black hole in the hyperboloidal framework}",
    eprint = "2412.21092",
    archivePrefix = "arXiv",
    primaryClass = "gr-qc",
    reportNumber = "ICTS-USTC/PCFT-24-58",
    doi = "10.1007/s11433-025-2714-7",
    journal = "Sci. China Phys. Mech. Astron.",
    volume = "68",
    number = "10",
    pages = "100411",
    year = "2025"
}

@article{Cao:2025qws,
    author = "Cao, Li-Ming and Ji, Ming-Fei and Wu, Liang-Bi and Zhou, Yu-Sen",
    title = "{Pseudospectrum and time-domain analysis of the EFT corrected black holes}",
    eprint = "2508.13894",
    archivePrefix = "arXiv",
    primaryClass = "gr-qc",
    reportNumber = "ICTS-USTC/PCFT-25-33",
    doi = "10.1103/jc1v-g6hc",
    journal = "Phys. Rev. D",
    volume = "112",
    number = "12",
    pages = "124022",
    year = "2025"
}

@article{Cardoso:2016rao,
    author = "Cardoso, Vitor and Franzin, Edgardo and Pani, Paolo",
    title = "{Is the gravitational-wave ringdown a probe of the event horizon?}",
    eprint = "1602.07309",
    archivePrefix = "arXiv",
    primaryClass = "gr-qc",
    doi = "10.1103/PhysRevLett.116.171101",
    journal = "Phys. Rev. Lett.",
    volume = "116",
    number = "17",
    pages = "171101",
    year = "2016",
    note = "[Erratum: Phys.Rev.Lett. 117, 089902 (2016)]"
}

@article{Cardoso:2019apo,
    author = "Cardoso, Vitor and Foit, Valentino F. and Kleban, Matthew",
    title = "{Gravitational wave echoes from black hole area quantization}",
    eprint = "1902.10164",
    archivePrefix = "arXiv",
    primaryClass = "hep-th",
    doi = "10.1088/1475-7516/2019/08/006",
    journal = "JCAP",
    volume = "08",
    pages = "006",
    year = "2019"
}

@article{Cardoso:2024mrw,
    author = "Cardoso, Vitor and Kastha, Shilpa and Panosso Macedo, Rodrigo",
    title = "{Physical significance of the black hole quasinormal mode spectra instability}",
    eprint = "2404.01374",
    archivePrefix = "arXiv",
    primaryClass = "gr-qc",
    doi = "10.1103/PhysRevD.110.024016",
    journal = "Phys. Rev. D",
    volume = "110",
    number = "2",
    pages = "024016",
    year = "2024"
}

@article{Chakraborty:2023zed,
    author = "Chakraborty, Sumanta and Maggio, Elisa and Silvestrini, Michela and Pani, Paolo",
    title = "{Dynamical tidal Love numbers of Kerr-like compact objects}",
    eprint = "2310.06023",
    archivePrefix = "arXiv",
    primaryClass = "gr-qc",
    doi = "10.1103/PhysRevD.110.084042",
    journal = "Phys. Rev. D",
    volume = "110",
    number = "8",
    pages = "084042",
    year = "2024"
}

@article{Chen:2024mon,
    author = "Chen, Jia-Ning and Wu, Liang-Bi and Guo, Zong-Kuan",
    title = "{The pseudospectrum and transient of Kaluza{\textendash}Klein black holes in Einstein{\textendash}Gauss{\textendash}Bonnet gravity}",
    eprint = "2407.03907",
    archivePrefix = "arXiv",
    primaryClass = "gr-qc",
    doi = "10.1088/1361-6382/ad89a1",
    journal = "Class. Quant. Grav.",
    volume = "41",
    number = "23",
    pages = "235015",
    year = "2024"
}

@article{Cheung:2021bol,
    author = "Cheung, Mark Ho-Yeuk and Destounis, Kyriakos and Macedo, Rodrigo Panosso and Berti, Emanuele and Cardoso, Vitor",
    title = "{Destabilizing the Fundamental Mode of Black Holes: The Elephant and the Flea}",
    eprint = "2111.05415",
    archivePrefix = "arXiv",
    primaryClass = "gr-qc",
    doi = "10.1103/PhysRevLett.128.111103",
    journal = "Phys. Rev. Lett.",
    volume = "128",
    number = "11",
    pages = "111103",
    year = "2022"
}

@article{Courty:2023rxk,
    author = "Courty, Aubin and Destounis, Kyriakos and Pani, Paolo",
    title = "{Spectral instability of quasinormal modes and strong cosmic censorship}",
    eprint = "2307.11155",
    archivePrefix = "arXiv",
    primaryClass = "gr-qc",
    doi = "10.1103/PhysRevD.108.104027",
    journal = "Phys. Rev. D",
    volume = "108",
    number = "10",
    pages = "104027",
    year = "2023"
}

@article{Cownden:2023dam,
    author = "Cownden, Bradley and Pantelidou, Christiana and Zilh{\~a}o, Miguel",
    title = "{The pseudospectra of black holes in AdS}",
    eprint = "2312.08352",
    archivePrefix = "arXiv",
    primaryClass = "gr-qc",
    doi = "10.1007/JHEP05(2024)202",
    journal = "JHEP",
    volume = "05",
    pages = "202",
    year = "2024"
}

@article{Daghigh:2020jyk,
    author = "Daghigh, Ramin G. and Green, Michael D. and Morey, Jodin C.",
    title = "{Significance of Black Hole Quasinormal Modes: A Closer Look}",
    eprint = "2002.07251",
    archivePrefix = "arXiv",
    primaryClass = "gr-qc",
    doi = "10.1103/PhysRevD.101.104009",
    journal = "Phys. Rev. D",
    volume = "101",
    number = "10",
    pages = "104009",
    year = "2020"
}

@article{Deppe:2024fdo,
    author = "Deppe, Nils and Heisenberg, Lavinia and Inchausp{\'e}, Henri and Kidder, Lawrence E. and Maibach, David and Ma, Sizheng and Moxon, Jordan and Nelli, Kyle C. and Throwe, William and Vu, Nils L.",
    title = "{Echoes from beyond: Detecting gravitational-wave quantum imprints with LISA}",
    eprint = "2411.05645",
    archivePrefix = "arXiv",
    primaryClass = "gr-qc",
    doi = "10.1103/k7jh-rhgw",
    journal = "Phys. Rev. D",
    volume = "111",
    number = "12",
    pages = "124035",
    year = "2025"
}

@article{Destounis:2021lum,
    author = "Destounis, Kyriakos and Macedo, Rodrigo Panosso and Berti, Emanuele and Cardoso, Vitor and Jaramillo, Jos{\'e} Luis",
    title = {{Pseudospectrum of Reissner-Nordstr{\"o}m black holes: Quasinormal mode instability and universality}},
    eprint = "2107.09673",
    archivePrefix = "arXiv",
    primaryClass = "gr-qc",
    doi = "10.1103/PhysRevD.104.084091",
    journal = "Phys. Rev. D",
    volume = "104",
    number = "8",
    pages = "084091",
    year = "2021"
}

@inproceedings{Destounis:2023ruj,
    author = "Destounis, Kyriakos and Duque, Francisco",
    title = "{Black-hole spectroscopy: quasinormal modes, ringdown stability and the pseudospectrum}",
    eprint = "2308.16227",
    archivePrefix = "arXiv",
    primaryClass = "gr-qc",
    doi = "10.1007/978-3-031-55098-0_6",
    month = "8",
    year = "2023"
}

@article{Destounis:2023nmb,
    author = "Destounis, Kyriakos and Boyanov, Valentin and Panosso Macedo, Rodrigo",
    title = "{Pseudospectrum of de Sitter black holes}",
    eprint = "2312.11630",
    archivePrefix = "arXiv",
    primaryClass = "gr-qc",
    doi = "10.1103/PhysRevD.109.044023",
    journal = "Phys. Rev. D",
    volume = "109",
    number = "4",
    pages = "044023",
    year = "2024"
}

@article{Destounis:2025dck,
    author = "Destounis, Kyriakos and Malato Corr{\^e}a, Mateus and Macedo, Caio F. B. and Panosso Macedo, Rodrigo",
    title = "{Spectral instability of horizonless compact objects within astrophysical environments: The ''exotic'' elephant and the flea}",
    eprint = "2509.16310",
    archivePrefix = "arXiv",
    primaryClass = "gr-qc",
    month = "9",
    year = "2025"
}

@article{Hu:2025efp,
    author = "Hu, Han-Wen and Lan, Chen and Guo, Zong-Kuan",
    title = "{Parity breaking reshapes black hole spectral dynamics}",
    eprint = "2511.21786",
    archivePrefix = "arXiv",
    primaryClass = "gr-qc",
    month = "11",
    year = "2025"
}

@article{Ianniccari:2024ysv,
    author = "Ianniccari, A. and Iovino, A. J. and Kehagias, A. and Pani, P. and Perna, G. and Perrone, D. and Riotto, A.",
    title = "{Deciphering the Instability of the Black Hole Ringdown Quasinormal Spectrum}",
    eprint = "2407.20144",
    archivePrefix = "arXiv",
    primaryClass = "gr-qc",
    doi = "10.1103/PhysRevLett.133.211401",
    journal = "Phys. Rev. Lett.",
    volume = "133",
    number = "21",
    pages = "211401",
    year = "2024"
}

@article{Jaramillo:2020tuu,
    author = "Jaramillo, Jos{\'e} Luis and Panosso Macedo, Rodrigo and Al Sheikh, Lamis",
    title = "{Pseudospectrum and Black Hole Quasinormal Mode Instability}",
    eprint = "2004.06434",
    archivePrefix = "arXiv",
    primaryClass = "gr-qc",
    doi = "10.1103/PhysRevX.11.031003",
    journal = "Phys. Rev. X",
    volume = "11",
    number = "3",
    pages = "031003",
    year = "2021"
}

@article{Jaramillo:2021tmt,
    author = "Jaramillo, Jos{\'e} Luis and Panosso Macedo, Rodrigo and Sheikh, Lamis Al",
    title = "{Gravitational Wave Signatures of Black Hole Quasinormal Mode Instability}",
    eprint = "2105.03451",
    archivePrefix = "arXiv",
    primaryClass = "gr-qc",
    doi = "10.1103/PhysRevLett.128.211102",
    journal = "Phys. Rev. Lett.",
    volume = "128",
    number = "21",
    pages = "211102",
    year = "2022"
}

@article{Kokkotas:1999bd,
    author = "Kokkotas, Kostas D. and Schmidt, Bernd G.",
    title = "{Quasinormal modes of stars and black holes}",
    eprint = "gr-qc/9909058",
    archivePrefix = "arXiv",
    doi = "10.12942/lrr-1999-2",
    journal = "Living Rev. Rel.",
    volume = "2",
    pages = "2",
    year = "1999"
}

@article{Konoplya:2011qq,
    author = "Konoplya, R. A. and Zhidenko, A.",
    title = "{Quasinormal modes of black holes: From astrophysics to string theory}",
    eprint = "1102.4014",
    archivePrefix = "arXiv",
    primaryClass = "gr-qc",
    doi = "10.1103/RevModPhys.83.793",
    journal = "Rev. Mod. Phys.",
    volume = "83",
    pages = "793--836",
    year = "2011"
}

@article{Konoplya:2022pbc,
    author = "Konoplya, R. A. and Zhidenko, A.",
    title = "{First few overtones probe the event horizon geometry}",
    eprint = "2209.00679",
    archivePrefix = "arXiv",
    primaryClass = "gr-qc",
    doi = "10.1016/j.jheap.2024.10.015",
    journal = "JHEAp",
    volume = "44",
    pages = "419--426",
    year = "2024"
}

@article{Liu:2021aqh,
    author = "Liu, Hang and Qian, Wei-Liang and Liu, Yunqi and Wu, Jian-Pin and Wang, Bin and Yue, Rui-Hong",
    title = "{Alternative mechanism for black hole echoes}",
    eprint = "2104.11912",
    archivePrefix = "arXiv",
    primaryClass = "gr-qc",
    doi = "10.1103/PhysRevD.104.044012",
    journal = "Phys. Rev. D",
    volume = "104",
    number = "4",
    pages = "044012",
    year = "2021"
}

@article{Mai:2025cva,
    author = "Mai, Zhan-Feng and Yang, Run-Qiu",
    title = "{Butterfly in Spacetime: Inherent Instabilities in Stable Black Holes}",
    eprint = "2506.07562",
    archivePrefix = "arXiv",
    primaryClass = "gr-qc",
    month = "6",
    year = "2025"
}

@article{MalatoCorrea:2025iuc,
    author = "Malato Corr{\^e}a, Mateus and Macedo, Caio F. B. and Panosso Macedo, Rodrigo and Oliveira, Leandro A.",
    title = "{Black hole spectral instabilities in the laboratory: Shallow water analog}",
    eprint = "2504.00107",
    archivePrefix = "arXiv",
    primaryClass = "gr-qc",
    doi = "10.1103/78ht-dn36",
    journal = "Phys. Rev. D",
    volume = "112",
    number = "2",
    pages = "024036",
    year = "2025"
}

@article{Mark:2017dnq,
    author = "Mark, Zachary and Zimmerman, Aaron and Du, Song Ming and Chen, Yanbei",
    title = "{A recipe for echoes from exotic compact objects}",
    eprint = "1706.06155",
    archivePrefix = "arXiv",
    primaryClass = "gr-qc",
    reportNumber = "LIGO-P1700145",
    doi = "10.1103/PhysRevD.96.084002",
    journal = "Phys. Rev. D",
    volume = "96",
    number = "8",
    pages = "084002",
    year = "2017"
}

@article{Nollert:1996rf,
    author = "Nollert, Hans-Peter",
    title = "{About the significance of quasinormal modes of black holes}",
    eprint = "gr-qc/9602032",
    archivePrefix = "arXiv",
    doi = "10.1103/PhysRevD.53.4397",
    journal = "Phys. Rev. D",
    volume = "53",
    pages = "4397--4402",
    year = "1996"
}

@article{Nollert:1998ys,
    author = "Nollert, Hans-Peter and Price, Richard H.",
    title = "{Quantifying excitations of quasinormal mode systems}",
    eprint = "gr-qc/9810074",
    archivePrefix = "arXiv",
    doi = "10.1063/1.532698",
    journal = "J. Math. Phys.",
    volume = "40",
    pages = "980--1010",
    year = "1999"
}

@article{Oshita:2019sat,
    author = "Oshita, Naritaka and Wang, Qingwen and Afshordi, Niayesh",
    title = "{On Reflectivity of Quantum Black Hole Horizons}",
    eprint = "1905.00464",
    archivePrefix = "arXiv",
    primaryClass = "hep-th",
    doi = "10.1088/1475-7516/2020/04/016",
    journal = "JCAP",
    volume = "04",
    pages = "016",
    year = "2020"
}

@article{Oshita:2025ibu,
    author = "Oshita, Naritaka and Berti, Emanuele and Cardoso, Vitor",
    title = "{Unstable Chords and Destructive Resonant Excitation of Black Hole Quasinormal Modes}",
    eprint = "2503.21276",
    archivePrefix = "arXiv",
    primaryClass = "gr-qc",
    reportNumber = "YITP-25-44, RIKEN-iTHEMS-Report-25",
    doi = "10.1103/ht2n-vvvh",
    journal = "Phys. Rev. Lett.",
    volume = "135",
    number = "3",
    pages = "031401",
    year = "2025"
}

@article{Qian:2020cnz,
    author = "Qian, Wei-Liang and Lin, Kai and Shao, Cai-Ying and Wang, Bin and Yue, Rui-Hong",
    title = "{Asymptotical quasinormal mode spectrum for piecewise approximate effective potential}",
    eprint = "2009.11627",
    archivePrefix = "arXiv",
    primaryClass = "gr-qc",
    doi = "10.1103/PhysRevD.103.024019",
    journal = "Phys. Rev. D",
    volume = "103",
    number = "2",
    pages = "024019",
    year = "2021"
}

@article{Shen:2025nsl,
    author = "Shen, Shui-Fa and Li, Guan-Ru and Daghigh, Ramin G. and Morey, Jodin C. and Green, Michael D. and Qian, Wei-Liang and Yue, Rui-Hong",
    title = "{Black hole spectral instability triggered by deterministic and random metric perturbations}",
    eprint = "2509.23372",
    archivePrefix = "arXiv",
    primaryClass = "gr-qc",
    month = "9",
    year = "2025"
}

@article{Solidoro:2024yxi,
    author = "Solidoro, Leonardo and Patrick, Sam and Weinfurtner, Silke and Gregory, Ruth",
    title = "{Origin of Quasinormal Modes in Semi-Open Systems}",
    eprint = "2406.11013",
    archivePrefix = "arXiv",
    primaryClass = "gr-qc",
    doi = "10.1103/7kjp-vrml",
    journal = "Phys. Rev. Lett.",
    volume = "135",
    number = "5",
    pages = "051401",
    year = "2025"
}

@book{trefethen2020spectra,
    author = {Trefethen, Lloyd and Embree, Mark},
    year = {2005},
    month = {01},
    pages = {},
    title = {Spectra and Pseudospectra: The Behavior of Nonnormal Matrices and Operators},
    isbn = {9780691119465},
    doi = {10.2307/j.ctvzxx9kj},
    publisher={Princeton university press}
}

@article{Wang:2019rcf,
    author = "Wang, Qingwen and Oshita, Naritaka and Afshordi, Niayesh",
    title = "{Echoes from Quantum Black Holes}",
    eprint = "1905.00446",
    archivePrefix = "arXiv",
    primaryClass = "gr-qc",
    doi = "10.1103/PhysRevD.101.024031",
    journal = "Phys. Rev. D",
    volume = "101",
    number = "2",
    pages = "024031",
    year = "2020"
}

@article{Wang:2025mxe,
    author = "Wang, Peng and Wu, Tianshu",
    title = "{Coexistence of Spectrally Stable and Unstable Modes in Black Hole Ringdowns}",
    eprint = "2510.12135",
    archivePrefix = "arXiv",
    primaryClass = "gr-qc",
    reportNumber = "CTP-SCU/20250017",
    month = "10",
    year = "2025"
}

@article{Yang:2024vor,
    author = "Yang, Yiqiu and Mai, Zhan-Feng and Yang, Run-Qiu and Shao, Lijing and Berti, Emanuele",
    title = "{Spectral instability of black holes: Relating the frequency domain to the time domain}",
    eprint = "2407.20131",
    archivePrefix = "arXiv",
    primaryClass = "gr-qc",
    doi = "10.1103/PhysRevD.110.084018",
    journal = "Phys. Rev. D",
    volume = "110",
    number = "8",
    pages = "084018",
    year = "2024"
}

@article{Bonelli:2022ten,
    author = "Bonelli, Giulio and Iossa, Cristoforo and Panea Lichtig, Daniel and Tanzini, Alessandro",
    title = "{Irregular Liouville Correlators and Connection Formulae for Heun Functions}",
    eprint = "2201.04491",
    archivePrefix = "arXiv",
    primaryClass = "hep-th",
    doi = "10.1007/s00220-022-04497-5",
    journal = "Commun. Math. Phys.",
    volume = "397",
    number = "2",
    pages = "635--727",
    year = "2023"
}

@article{Suzuki:1998vy,
    author = "Suzuki, Hisao and Takasugi, Eiichi and Umetsu, Hiroshi",
    title = "{Perturbations of Kerr-de Sitter black hole and Heun's equations}",
    eprint = "gr-qc/9805064",
    archivePrefix = "arXiv",
    reportNumber = "EPHOU-98-005, OU-HET-296",
    doi = "10.1143/PTP.100.491",
    journal = "Prog. Theor. Phys.",
    volume = "100",
    pages = "491--505",
    year = "1998"
}

@article{Hatsuda:2020sbn,
    author = "Hatsuda, Yasuyuki",
    title = "{Quasinormal modes of Kerr-de Sitter black holes via the Heun function}",
    eprint = "2006.08957",
    archivePrefix = "arXiv",
    primaryClass = "gr-qc",
    reportNumber = "RUP-20-19",
    doi = "10.1088/1361-6382/abc82e",
    journal = "Class. Quant. Grav.",
    volume = "38",
    number = "2",
    pages = "025015",
    year = "2020"
}

@article{Suzuki:1999tsi,
  author = {Suzuki, Hisao and Takasugi, Eiichi and Umetsu, Hiroshi},
  title = {Analytic Solutions of Teukolsky Equation in Kerr-de Sitter and Kerr-Newman-de Sitter Geometries},
  journal = {Prog. Theor. Phys.},
  volume = {102},
  pages = {253--272},
  year = {1999},
  doi = {10.1143/PTP.102.253},
  eprint = {gr-qc/9905040},
  archivePrefix = {arXiv}
}

@article{Suzuki:2000abs,
  author = {Suzuki, Hisao and Takasugi, Eiichi and Umetsu, Hiroshi},
  title = {Absorption Rate of the Kerr-de Sitter Black Hole and the Kerr-Newman-de Sitter Black Hole},
  journal = {Prog. Theor. Phys.},
  volume = {103},
  pages = {723--731},
  year = {2000},
  doi = {10.1143/PTP.103.723},
  eprint = {gr-qc/9911079},
  archivePrefix = {arXiv}
}

@article{Chen:2021tfc,
  author = {Chen, Baoyi and Wang, Qingwen and Chen, Yanbei},
  title = {Tidal response and near-horizon boundary conditions for spinning exotic compact objects},
  journal = {Phys. Rev. D},
  volume = {103},
  pages = {104054},
  year = {2021},
  doi = {10.1103/PhysRevD.103.104054},
  eprint = {2012.10842},
  archivePrefix = {arXiv},
  primaryClass = {gr-qc}
}

@article{Nakano:2017howls,
  author = {Nakano, Hiroyuki and Sago, Norichika and Tagoshi, Hideyuki and Tanaka, Takahiro},
  title = {Black hole ringdown echoes and howls},
  journal = {Prog. Theor. Exp. Phys.},
  volume = {2017},
  pages = {071E01},
  year = {2017},
  doi = {10.1093/ptep/ptx093}
}

@article{Novaes:2019ynt,
    author = "Novaes, F{\'a}bio and Marinho, C{\'a}ssio I. S. and Lencs{\'e}s, M{\'a}t{\'e} and Casals, Marc",
    title = "{Kerr-de Sitter quasinormal modes via accessory parameter expansion}",
    journal = "JHEP",
    volume = "05",
    pages = "033",
    year = "2019",
    doi = "10.1007/JHEP05(2019)033"
}

@article{Maggio:2017ivp,
    author = "Maggio, Elisa and Pani, Paolo and Ferrari, Valeria",
    title = "{Exotic compact objects and how to quench their ergoregion instability}",
    journal = "Phys. Rev. D",
    volume = "96",
    pages = "104047",
    year = "2017",
    eprint = "1703.03696",
    archivePrefix = "arXiv",
    primaryClass = "gr-qc",
    doi = "10.1103/PhysRevD.96.104047"
}

@article{Maggio:2019qhr,
    author = "Maggio, Elisa and Cardoso, Vitor and Dolan, Sam R. and Pani, Paolo",
    title = "{Ergoregion instability of exotic compact objects: Electromagnetic and gravitational perturbations and the role of absorption}",
    journal = "Phys. Rev. D",
    volume = "99",
    pages = "064007",
    year = "2019",
    doi = "10.1103/PhysRevD.99.064007"
}

@article{Leaver:1985ax,
    author = "Leaver, E. W.",
    title = "{An analytic representation for the quasi-normal modes of Kerr black holes}",
    journal = "Proc. Roy. Soc. Lond. A",
    volume = "402",
    pages = "285--298",
    year = "1985",
    doi = "10.1098/rspa.1985.0119"
}

@article{Yang:2013qnm,
  author = {Yang, Huan and Zimmerman, Aaron and Zengino{\u{g}}lu, An{\i}l and Zhang, Fan and Berti, Emanuele and Chen, Yanbei},
  title = {Quasinormal modes of nearly extremal Kerr spacetimes: Spectrum bifurcation and power-law ringdown},
  journal = {Phys. Rev. D},
  volume = {88},
  pages = {044047},
  year = {2013},
  doi = {10.1103/PhysRevD.88.044047},
  eprint = {1307.8086},
  archivePrefix = {arXiv},
  primaryClass = {gr-qc}
}

@article{Teukolsky:1973,
    author = {Teukolsky, Saul A.},
    title = {Perturbations of a Rotating Black Hole. I. Fundamental Equations for Gravitational, Electromagnetic, and Neutrino-Field Perturbations},
    journal = {Astrophys. J.},
    volume = {185},
    pages = {635--647},
    year = {1973},
    doi = {10.1086/152444}
}

@article{Khanal:1983,
    author = {Khanal, Udayaraj},
    title = {Rotating black hole in asymptotic de Sitter space: Perturbation of the space-time with spin fields},
    journal = {Phys. Rev. D},
    volume = {28},
    number = {6},
    pages = {1291--1297},
    year = {1983},
    doi = {10.1103/PhysRevD.28.1291}
}

@article{Chambers:1994,
    author = {Chambers, Chris M. and Moss, Ian G.},
    title = {Stability of the Cauchy horizon in Kerr-de Sitter space-times},
    journal = {Class. Quant. Grav.},
    volume = {11},
    pages = {1035--1054},
    year = {1994},
    doi = {10.1088/0264-9381/11/4/019},
    eprint = {gr-qc/9404015},
    archivePrefix = {arXiv}
}

@article{Wu:2025wbp,
    author = "Wu, Liang-Bi and Xie, Libo and Cao, Li-Ming and Ji, Ming-Fei and Zhou, Yu-Sen",
    title = "{Quasinormal modes of Schwarzschild-de Sitter black holes in semi-open systems}",
    eprint = "2512.06903",
    archivePrefix = "arXiv",
    primaryClass = "gr-qc",
    doi = "10.1007/s11433-025-2912-y",
    journal = "Sci. China Phys. Mech. Astron.",
    volume = "69",
    number = "4",
    pages = "240415",
    year = "2026"
}

@article{Zhou:2025xdo,
    author = "Zhou, Yu-Sen and Ji, Ming-Fei and Wu, Liang-Bi and Cao, Li-Ming",
    title = "{Pseudospectrum and (in)stability of black hole total transmission modes}",
    eprint = "2512.16372",
    archivePrefix = "arXiv",
    primaryClass = "gr-qc",
    reportNumber = "ICTS-USTC/PCFT-25-63",
    doi = "10.1103/plvy-f3mf",
    journal = "Phys. Rev. D",
    volume = "113",
    number = "12",
    pages = "124042",
    year = "2026"
}

@article{Wu:2026xeg,
    author = "Wu, Liang-Bi and Zhou, Yu-Sen and Ji, Ming-Fei and Liu, Xia-Yuan and Xie, Libo and Cao, Li-Ming",
    title = "{The structured pseudospectrum of Schwarzschild black hole}",
    eprint = "2611.00853",
    archivePrefix = "arXiv",
    primaryClass = "gr-qc",
    reportNumber = "ICTS-USTC/PCFT-26-79",
    month = "10",
    year = "2026"
}

@article{Zhang:2026cjq,
    author = "Zhang, Hongsheng and Guo, Zong-Kuan and Cai, Rong-Gen",
    title = "{The pseudospectrum for gravitational perturbations in Kerr spacetimes}",
    eprint = "2610.06954",
    archivePrefix = "arXiv",
    primaryClass = "gr-qc",
    month = "10",
    year = "2026"
}
\bibliographystyle{apsrev4-1}

\end{document}